\documentclass{WileyMSP-template}
\usepackage[margin=0.5in]{geometry}
\usepackage{graphicx}
\usepackage{caption}
\usepackage{fancyhdr}
\usepackage{gensymb}
\usepackage{rotating}
\usepackage{ragged2e}
\usepackage{times,mathptmx}
\usepackage{amsmath} 
\usepackage[T1]{fontenc}
\usepackage{lmodern}
\begin{document}

\pagestyle{fancy}
\rhead{\includegraphics[width=2.5cm]{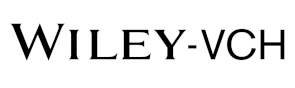}}

\title{Structural phase stability, electronic structure, magnetic properties and chemical bonding analysis of transition metal ammine borohydrides with amphoteric hydrogen for hydrogen storage}

\maketitle


\author{S Kiruthika and}
\author{P Ravindran*}



\begin{affiliations}
Prof. P. Ravindran\\
Department of Physics, School of Basics and Applied Science, Central University of Tamil Nadu, Thiruvarur, India. 
Simulation Center for Atomic and Nanoscale MATerials (SCANMAT), Central University of Tamil Nadu, Thiruvarur, India
Department of Chemistry, University of Oslo, Box 1033 Blindern, N0315, Norway.
Email Address:raviphy@cutn.ac.in

S Kiruthika\\
Department of Physics, School of Basics and Applied Science, Central University of Tamil Nadu, Thiruvarur, India.\\
\end{affiliations}


\keywords{Amphoteric hydrogen, Hydrogen storage, van der Waals interaction, Magnetic Properties, Computational.}

\begin{abstract}
\justifying
Usually the ions of a particular element in solids are in positive or negative oxidation states, depending upon the chemical environment. It is highly unusual for an atom having both positive as well as negative oxidation states simultaneously within the same structural framework in a particular compound. Our structural and chemical bonding analyses show that the hydrogen ions in the transition metal ammine borohydrides (TMABHs) with the chemical formula $M$(BH$_4$)$_2$(NH$_3$)$_2$ ($M$ = Sc, Ti, V, Cr, Mn, Fe, Co, Ni, Cu and Zn) have both $+$ and $-$ oxidation states. Based on the structural optimization using force and stress minimization we have calculated the ground state crystal structure and equilibrium structural parameters for TMABHs. We have found that the ground state structures for experimentally known systems may reliably be predicted only when we include van der Waals correction to the spin-polarised total energy calculation for these compounds. Our detailed analyses show that the hydrogen present in these compounds have amphoteric behavior with hydrogen closer to boron is in negative oxidation state and that closer to nitrogen is in the positive oxidation state. The spin-polarized van der Waals interaction included calculation show that all these materials except M = Zn are having finite magnetic moment at the transitions metal site with the anti$-$ferromagnetic ordering as ground state. The localised magnetic moment similar to those present in molecular magnet with large band gap value indicating that these transparent magnets may find application in novel devices. Our nudged elastic band calculation show that the migration barrier for ammonia diffusion in these materials are more than 1 eV and hence ammonia poisoning during hydrogen release is unlikely and hence these compounds may be used for energy storage applications since they have high weight percentage of hydrogen. The chemical bonding interactions between constituents were characterised using various bonding analyses tools. Due to the presence of finite covalent bonding between the constituents in TMABHs the oxidation state of hydrogen is non$-$integer value. The confirmation of the presence of amphoteric behaviour of hydrogen in TMABHs has implication in designing volume efficient hydrides for hydrogen storage applications.
\end{abstract}


\section{Introduction}
\justifying
Hydrogen is one of the ideal and promising environmental benign energy carriers. However, developing safe and efficient hydrogen storage materials for fuel cell applications is a key challenge. The complex hydrides are promising to be an excellent hydrogen storage materials compared to other hydrides. \cite{rusman2016review, hong2013hydrogen, jain2007hydrogenation, jain2010hydrogen} The complex hydrides are categorized into alanates, borohydrides, amides, etc. For the past few years, metal based borohydrides are considered as potential hydrogen storage materials owing to their large gravimetric hydrogen densities, low toxicity and low volatility compared to other complex hydrides.\cite{nickels2008tuning} Even though the metal based borohydrides are considered as the promising materials for the hydrogen storage applications, the key challenge to use them for practical applications is the slow kinetics and poor reversibility added with high decomposition temperature. ~\cite{bogdanovic1997ti, schuth2004light, li2011recent, chen2002interaction} Let's have a concise description about the metal based borohydrides.

\paragraph{}In 1939 Schlesinger \textit{et al.}~\cite{schlesinger1939volatile} were debutant prepared aluminium borohydride, Al(BH$_{4}$)$_{3}$, the first complex metal based borohydride. In continuum with Schlesinger, Brown and Brug has resolved the crystal structure of Be(BH$_{4}$)$_{2}$, and LiBH$_{4}$.~\cite{burg1940metallo, schlesinger1940metallo, schlesinger1940metalloIII}. The NaBH$_{4}$ was the first metal borohydride which was used as a hydrogen carrier during the second world war. In 1991 Dain \textit{et al.}~\cite{dain1991molecular} characterized the titanium metal based  borohydride Ti(BH$_{4}$)$_{3}$ in gaseous phase, which was the first identified transition metal borohydride. Radovan \textit{et al.}~\cite{cerny2009first}have resolved the crystal structure of Mn(BH$_{4}$)$_{2}$ in 2009. Later, Z{\"u}ttel \textit{et al.}~\cite{zuttel2003libh4, zuttel2003hydrogen} have studied the properties of LiBH$_{4}$ and shown that SiO$_2$ can be used as a catalyst to desorb hydrogen from LiBH$_{4}$ and 13.5 mass\% of hydrogen were liberated at 200 $^\circ$C. Vajo \textit{et al.}~\cite{vajo2005reversible, mauron2008stability} have studied the destabilization properties of LiBH$_{4}$ for reversible hydrogen storage using MgH$_{2}$ as a destabilizing additive. However, the kinetics for this system were too slow for direct measurements of hydrogen liberation at 225 $^{\circ}$C temperatures. In order to understand the thermal stability of the metal based borohydrides Nakamori \textit{et al.}~\cite{nakamori2007thermodynamical} synthesized and studied the desorption properties of the metal$-$borohydrides $M$(BH$_{4}$)$_{n}$ ($M$ = Ca, Sc, Ti, V, Cr, Mn, Zn and Al; n = 2$-$4). The results from their study indicate that the hydrogen desorption temperature, T$_{d}$, of $M$(BH$_{4}$)$_{n}$ correlate well with the Pauling electronegativity ($\chi_{P}$) value for $M$; that is, T$_{d}$ decreases with increasing value of $\chi_{P}$. So, a wide variety of metal based borohydrides were investigated and their properties were studied in detail. But the challenges to use them for practical hydrogen storage applications remain the same. Eventually, the researchers decided to have bimetallic borohydrides, in order to attain the minimum requirement to store the hydrogen efficiently.

\paragraph{} Both LiSc(BH$_{4}$)$_{4}$ and NaSc(BH$_{4}$)$_{4}$ were synthesised and their crystal structures were resolved in a decade ago.~\cite{hagemann2008lisc, cerny2009nasc} The crystal structure of LiSc(BH$_{4}$)$_{4}$ and NaSc(BH$_{4}$)$_{4}$ show disorderedness with the alkali metal and Sc(BH$_{4}$)$^{-}_{4}$ complex. The local structure of the Sc(BH$_{4}$)$^{-}_{4}$ complex is refined as a distorted form of the theoretical structure. It is found that the Li ions in the above system are found to be disordered along the z axis. The crystal structure of NaSc(BH$_{6}$)$_{4}$ is slightly deformed by trigonal Na$_{6}$ prisms. Considering the thermal stability of LiSc(BH$_{4}$)$_{4}$ and NaSc(BH$_{4}$)$_{4}$, Radovan \textit{et al.}~\cite{ cerny2009nasc} reported that the scandium boride ScB$_{x}$ may form as a decomposition product. This means that boron is stabilized in the solid dehydrogenated phase and hence, rehydrogenation of the decomposition products might be possible.

\paragraph{} Subsequently, Ravnsb{\ae}k\textit{et al.}~\cite{ravnsbaek2009series} have synthesized a new series of A$-$Zn$-$BH$_{4}$ borohydrides(A = Li,Na), via LiZn$_{2}$(BH$_{4}$)$_{5}$, NaZn$_{2}$(BH$_{4}$)$_{5}$, and NaZn(BH$_{4}$)$_{3}$. They have done a detailed study on the structural, physical, and chemical characterization of these Zn$-$based borohydrides. The Pauling electronegativity of zinc is higher than that of alkali metals and this difference may contribute to the lower stability of Zn based borohydrides. \cite{hwang2008nmr} In the Zn$-$rich (AZn$_{2}$(BH$_{4}$)$_{5}$) system, Zn and BH$_{4}$ units are strongly associated with isolated [Zn$_{2}$(BH$_{4}$)$_{5}$]$^{-}$ anions and A$^{+}$ as a counter cations. Similarly, in LiSc(BH$_{4}$)$_{4}$, the [Sc(BH$_{4}$)$ _{4} $]$^{-}$ unit is a isolated anion complex. \cite{hagemann2008lisc} This arrangement stabilizes this phase thereby enabling its synthesis at room temperature. Hence, Ravnsb{\ae}k \textit{et al.} suggested that the variation in the ratio between the alkali metal and the transition metal can tune the hydrogen storage properties in alkali metal$-$transition metal$-$BH$_{4}$ systems. However, it is a daunting task for the researchers to accomplish the key challenges such as slow kinetics, poor reversibility and high decomposition temperature exhibit in metal based borohydrides; even though we have a vast variety of such hydrides.

\paragraph{}
It may be noted that non-metal based borohydrides such as ammonia borane (AB) has been scrutinized for the hydrogen storage applications. \cite{karkamkar2007recent, chandra2007room} Ammonia borane (BH$_{3}$NH$_{3}$) is a promising hydrogen storage material because of its high gravimetric (19.6 wt\% H$_{2}$) and volumetric hydrogen density and moderate decomposition temperature. ~\cite{keaton2007base, gutowska2005nanoscaffold} It has been widely recognized as a promising candidate for hydrogen storage medium in mobile applications. The ammonia borane solid is stable at room temperature with a melting point of 383$-$388 K. During the last decade, thermal and catalyst induced dehydrogenation of AB has been intensively investigated.\cite{kumar2019solid} The thermal decomposition investigations show that the mass of the hydrogen released below 385 K is about 6.5 wt\% only through exothermic process and it is found that the variation of pressure does not significantly influence the reaction enthalpy and hydrogen release.\cite{zhong2012first, wolf2000calorimetric, baitalow2006thermal} The formation of AB is originated from different value of electronegativity of B and N. The electronegative N takes electrons from its three neighbouring H atoms leaving a partial positive charge at its neighbouring hydrogen sites. On the other hand, the electropositive boron gives electrons away to its three neighbouring H atoms, \cite{ghellab2019electronic} leaving a partial negative charge at these hydrogen sites. So, this solid is held together due to the Coulombic attraction between the opposite charges on the hydrogen atoms or dihydrogen bonds.\cite{hess2008spectroscopic, wolf1998thermochemical, west2009band} Using the density functional theory Miranda \textit{et al.}~\cite{miranda2007ab} have characterized the ammonia$-$borane complexes and found that the zero$-$point energy changes several H$_{2}$ release reactions from endothermic to exothermic. Both the ammonia$-$borane  polymer and borazine$-$cyclotriborazane show a strong exothermic decomposition character ($\approx$ $-$10 kcal mol$^{-1}$) implying that rehydrogenation may be difficult to achieve at moderate H$_{2}$ pressures and temperatures.

\paragraph{} Ammine metal borohydrides (AMBHs) possess the properties of both metal borohydrides as well as ammonia borane. Hence, these compounds exhibit gripping properties like high hydrogen capacity and favourable dehydrogenation properties. Soloveichik \textit{et al.} \cite{soloveichik2008ammine} described the synthesis, structural characterization, and hydrogen release properties of Mg(BH$_{4}$)$_{2}$ $\cdot$ 2NH$_{3}$. The relatively low hydrogen desorption temperature and high hydrogen capacity make Mg(BH$_{4}$)$_{2}$ $\cdot$ 2NH$_{3}$ a compelling candidate for hydrogen storage application competitive with ammonia borane. However, this material also suffers from the same problem possesed by borane and ammine complexes such as hydrogen purity. Jepsen \textit{et al.}~\cite{jepsen2015ammine} have introduced a series of solvent$-$ and halide$-$free alkali metal ammine borohydrides Sr(NH$_{3}$)$_{n}$(BH$_{4}$)$_{2}$ (n=1, 2, and 4) and Ca(NH$_{3}$)$_{n}$(BH$_{4} $)$_{2}$ (n=1, 2, 4, and 6) and investigated their thermo$-$decomposition behaviour. They found that these series of compounds release NH$_{3}$ gas upon thermal treatment if the partial pressure of ammonia is low. Finally, Jepsen \textit{et al.} revealed that the strength of the dihydrogen bonds, the crystal structure of these compounds, and the NH$_{3}$/BH$_{4}$ $^{-}$ ratio for Ca/Sr(NH$_{3}$)$_{n} $(BH$_{4} $)$_{m}$ have little influence on the composition of the released gases. Xiaowei\textit{et al.}~\cite{chen2012improved} reported a new combined system of Ca(BH$_{4}$)$_{2}$ $\cdot$ nNH$_{3}$ (n = 1, 2, and 4) and Mg(BH$ _{4}$)$_{2}$ complexes. Interestingly, the obtained composite exhibits a significant mutual dehydrogenation improvement compared with the pure Mg(BH$_{4}$)$_{2}$ or Ca(BH$_{4}$)$_{2}$ $\cdot$ nNH$_{3}$.

\paragraph{}
Yuan \textit{et al.}~\cite{yuan2012complex} have synthesized and studied the hydrogen storage properties of Ti(BH$_{4}$)$_{3}$ $\cdot$ 5NH$_{3}$, Li$_{2}$Ti(BH$_{4}$)$_{5}$ $\cdot$ 5NH$_{3}$, and Ti(BH$_{4} $)$ _{3} $ $\cdot$ 3NH$_{3}$. They found that Li$_{2}$Ti(BH$_{4}$)$_{5}$ $\cdot$ 5NH$_{3}$ and Ti(BH$_{4}$)$_{3}$ $\cdot$ 3NH$_{3}$ release $\sim$ 15.8 wt \% and 14 wt \% pure hydrogen, respectively below 300 $^{\circ}$C and over 9 wt \% pure hydrogen at a constant temperature of 100 $^{\circ}$C in both compounds, which meets the practical requirements for on$-$board hydrogen storage applications. These favourable dehydrogenation properties and potential regeneration ability make transitions metal ammine borohydrides very promising candidates as hydrogen storage materials. Roedern \textit{et al.}~\cite{roedern2015ammine} have synthesized [Fe(NH$_{3}$)$_{6}$](BH$_{4}$)$_{2}$ as well as [Co(NH$_{3}$)$_{6}$](BH$_{4}$)$_{2}$ and from that ammine iron borohydride$-$ammonia borane composites i.e. Fe(NH$_{3}$)$_{6}$(BH$_{4}$)$_{2}$ $\cdot$ nNH$_{3}$BH$_{3}$, where n = 2, 4, 6 series is prepared and analysed its gas release. The two components Fe(NH$_{3}$)$_{6}$(BH$_{4}$)$_{2}$ and NH$_{3}$BH$_{3}$ appear to decompose individually and hence the amount of released hydrogen content was not improved.

\paragraph{} The decomposition of metal based borohydrides attain in the range of 300 $^{\circ}$C and above this temperature one could expect the release of diborane B$_{2}$H$_{6}$. For example, LiBH$_{4}$ changes into an intermediate compound accompanying the release of approximately 11 wt\% of hydrogen at 430$-$460 $^{\circ}$C.~\cite{orimo2006experimental} Similarly, sodium borohydrides also  has a decomposition temperature of T$_{dec}$ = 534 $\pm$10 $^{\circ}$C at 1 bar for H$_{2}$ release~\cite{paskevicius2013eutectic, martelli2010stability}. However, the hydrogen start to release around T $\sim$ 500 $^{\circ}$C itself. Even though magnesium borohydride (Mg(BH$_{4}$)$_{2}$) received a significant interest as a possible hydrogen storage material, the hydrogen release occurs in three step process and the decomposition temperatures range from 265 to 400 $^{\circ}$C.~\cite{paskevicius2012situ, yan2015role} From the analysis of the thermodynamical properties of ammine metal borohydrides such as Al(BH$_{4}$)$_{3}$ $\cdot$ 6NH$_{3}$, Li$_{2}$Al(BH$_{4}$)$_{5}$ $\cdot$ 6NH$_{3}$ and Zn(BH$_{4}$)$_{2}$ $\cdot$ 2NH$_{3}$ it is found that the H$_{2}$ release with the trace of NH$_{3}$ is happening around the temperature range 115 to 170 $^{\circ}$. The ammine metal borohydrides with metal having low electronegativities ($\chi_{P}$ < 1.6) often display lower temperature for gas release when compared to that of the corresponding  pure metal borohydrides. On the other hand, ammine metal borohydrides having metal with high electronegativities ($\chi_{P}$ > 1.6) are often more thermally stable than their respective metal borohydride.~\cite{welchman2017decomposition, jepsen2015tailoring} Hence, the above circumstance make us to design new hydrogen storage materials based on transition metal ammine borohydrides where the transition metals have moderate electronegativity value compared with alkali metals so that hydrogen release will happen at reasonable temperature.

\paragraph{}
In most of the hydrogen storage materials hydrogen is in negative oxidation state and less than 10\% of the known hydrides alone hydrogen is in positive oxidation state. Further, if the hydrogen is in negative or positive oxidation states in solids, due to Coulombic repulsion of same charged hydrogen ions they are present with the inter$-$atomic distance 2~\AA\, or more.~\cite{switendick1979bandstructure} This limits the volumetric density of hydrogen in solids. If one can identify hydrogen storage materials where hydrogen is in both negative as well as positive oxidation states simultaneously, then due to Coulomb attraction of oppositely charged hydrogen reduce H$-$H separation and as a consequence of that, the gravimetric as well as the volumetric density of hydrogen in such systems can be increased. Designing such systems will pave the way to identify potential hydrogen storage materials. We have recently shown that it is possible to accommodate hydrogen in both positive and negative oxidation states simultaneously within the same structural frame work in systems such as (BH$_{3}$NH$_{3}$)$_{2}$ and (NH$_{4}$)$_{2}$(B$_{12}$H$_{12}$).~\cite{kiruthika2019amphoteric}

As mentioned above, compared with AB, the AMBHs have advantages hydrogen storage properties and we have already discussed about the chemical bonding nature of alkaline$-$earth ammine borohydrides(AMABHs).~\cite{kiruthika2018chemical} In the present study we are very much interested to investigate transition metal ammine borohydrides (TMABHs) using the density functional theory calculations to understand their structural stability, electronic structure, magnetic properties, chemical bonding analyses and decomposition mechanism. From the accurate van der Waals functional included spin$-$polarised electronic structure calculations, the ground state crystal structures and the equilibrium structural parameters were reproduced for experimentally known TMABHs such as Mn/Zn(BH$_4$)$_2$(NH$_3$)$_2$ and predicted the structural properties for the rest of the compounds in the 3\textit{d} transition metal based  TMABHs $M$(BH$_4$)$_2$(NH$_3$)$_2$ where ($M$ = Sc, Ti, V, Cr, Mn, Fe, Co, Ni, Cu and Zn). To substantiate the presence of amphoteric hydrogens in transition metal ammine borohydrides, we have calculated various chemical bonding analyses that will be discussed in section~\ref{cba}. In principle, most of the 3\textit{d} transition metal ions have magnetic behaviour. Moreover, TMABHs are have molecular$-$like transition metal structural sub units and hence studying their magnetic properties is of high current interest to understand the magnetic properties of molecular magnets. In order to substantiate the magnetic properties in TMABHs systems we have performed the spin- polarised band structure calculation where discussed in the section~\ref{Mag}. Also, the nudged elastic band (NEB) method calculations were performed to elucidate the decomposition mechanism (see section~\ref{DM}) of TMABHs.

\section{Result and Discussions}
\subsection{Structure description} \label{stdsp}
\begin{figure*}
\centering
 \includegraphics[width=\linewidth]{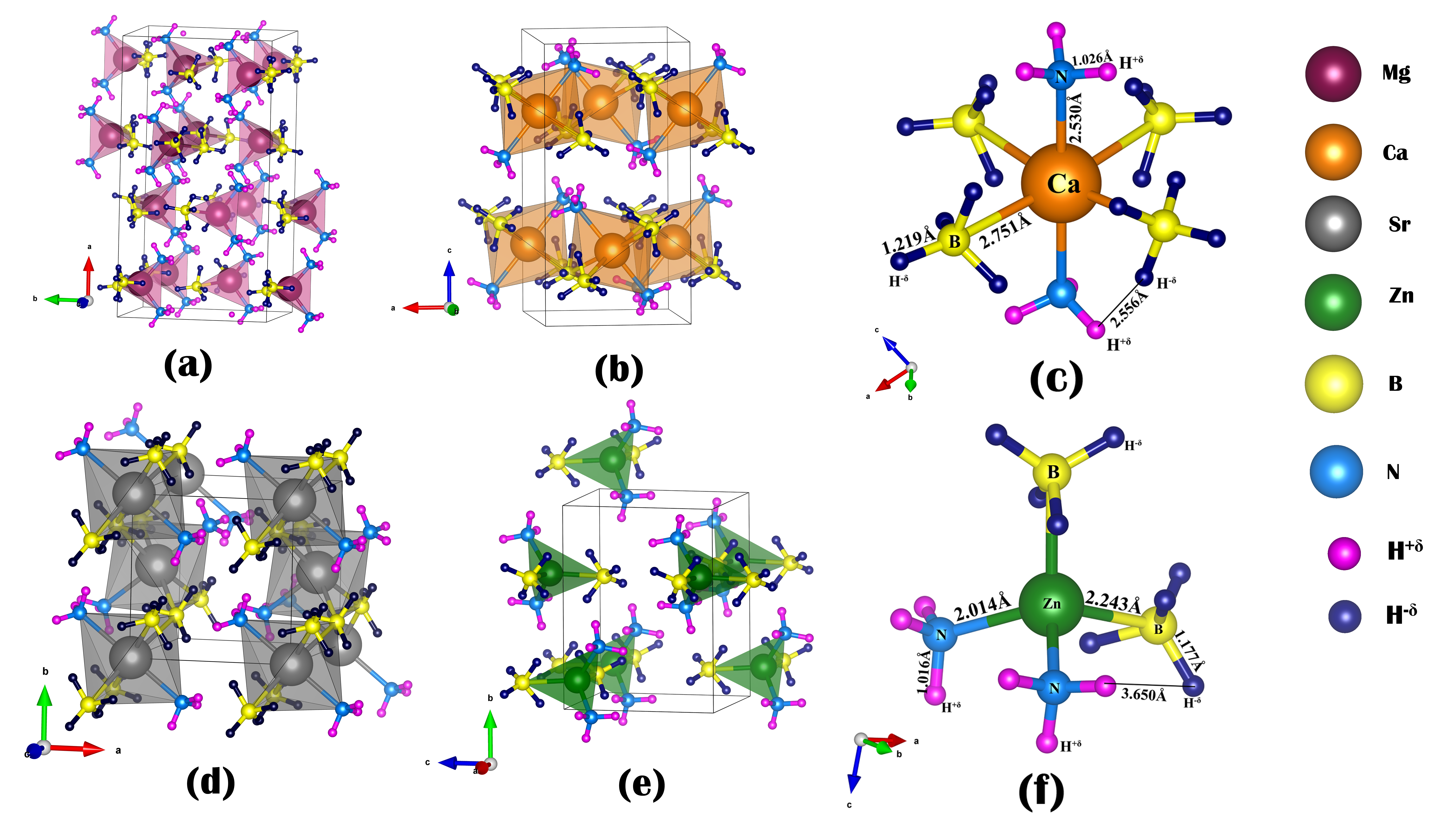}
  \caption{The crystal structures of $M$(BH$_4$)$_2$(NH$_3$)$_2$ ($M$ = Mg, Ca, Sr, or Zn)  are given in (a), (b), (d), and (e), respectively. Among these compounds $M$ = Ca and Sr are having octahedrally coordinated BH$_{4}$ and NH$_{3}$ structural sub-units and $M$ = Mg and Zn are having tetrahedrally coordinated BH$_{4}$ and NH$_{3}$ structural sub-units. Hence, the  structural complex of $M$ = Ca and Zn with bonds and their bond lengths are displayed in figure (c) and (f) as representative systems. The hydrogen with positive and negative oxidation states are depicted in rose and dark blue colour, respectively.}
   \label{cryst}
\end{figure*}

The crystal structures of $M$(BH$_4$)$_2$(NH$_3$)$_2$ ($M$ = Mg, Ca, Sr, Mn, and Zn) were identified experimentally and are illustrated in Fig.~\ref{cryst}. Among these compounds Mg/Mn(BH$_4$)$_2$(NH$_3$)$_2$,~\cite{soloveichik2008ammine, jepsen2015tailoring} Ca(BH$_4$)$_2$(NH$_3$)$_2$,~\cite{jepsen2015ammine} and Sr(BH$_4$)$_2$(NH$_3$)$_2$,~\cite{jepsen2015ammine} 
are stabilizing in the orthorhombic structure with space groups Pcab, Pbcn, and Pnc2, as depicted in Fig.~\ref{cryst} (a), (b), and (d), respectively. Whereas, Zn(BH$_4$)$_2$(NH$_3$)$_2$,~\cite{gu2012structure} has a monoclinic structure with space group P12$_{1}$1 as shown in Fig.~\ref{cryst} (e). The structures with space group Pbcn and Pnc2 have the structural arrangements such that the metal ions are octahedrally coordinated by four bridging (BH$_4$)$^-$ groups in the plane and two NH$_3$ groups axially over the plane.~\cite{Cotton1999} On the other hand, the metal ions are tetrahedrally coordinated by two bridging (BH$_4$)$^-$ groups in the plane and two NH$_3$ groups axially over the plane for Pcab and P12$_{1}$1 phases. Structural analysis of these compounds show that hydrogen atoms bonded to boron are expected to have negative oxidation state (H$^{-\delta}$). Similarly, the hydrogen atoms bonded to  nitrogen are expected to have positive oxidation states (H$^{+\delta}$). From the computationally optimized crystal structure using the optPBE$-$vdW functional, the estimated bond distance between B$-$ H$^{-\delta}$ is around 1.2~\AA ~whereas, that between N$-$H $^{+\delta}$ is around 1.0~\AA~(see Table~\ref{tbl:3}). The bond distance between H$^{-\delta}$ $-$ H$^{+\delta}$ in these compounds are ranges from 1.9 to 2.5~\AA ~for all the compounds considered in the present study.

\begin{figure}
\centering
 \includegraphics[scale = 0.5]{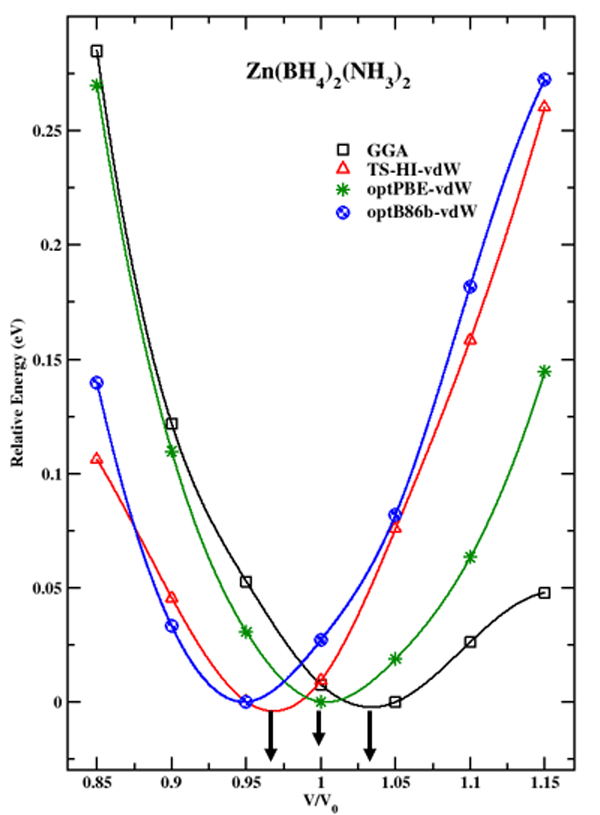}
 \caption{The total energy vs volume curves for Zn(BH$_4$)$_2$(NH$_3$)$_2$ obtained from \textit{ab$-$inito} calculation using GGA, TS$-$HI$-$vdW, optPBE$-$vdW, and optB86b$-$vdW functionals. The relative energy is obtained by scaling the total energy such that it will be zero for the equilibrium volume for each cases.}
 \label{znvol}
\end{figure}

In order to estimate the equilibrium structural parameters we have optimized the atomic coordinates and unit cell parameters globally using stress as well as force minimization with various exchange correlation functionals such as GGA, optPBE$-$vdW, optB86b$-$vdW, and TS$-$HI$-$vdW. The optimized equilibrium structural parameters are obtained by varying the unit cell volume between $-$15\% and 15\% from the experimental equilibrium volume with a step of 5\% and relaxing all atomic coordinates and shape of the unit cell globally for each volume. The resulting total energy vs. volume relationship for Zn(BH$_{4}$)$_{2}$(NH$_{3}$)$_{2}$ is illustrated in Fig.~\ref{znvol}. From these studies the calculated equilibrium cell volume, atomic coordinates, and unit cell dimensions are listed in Table~\ref{tbl:1} along with the available experimental~\cite{gu2012structure} as well as theoretical values obtained from optB86b$-$vdW functional earlier \cite{chen2018decomposition}.

It is evident from Fig.~\ref{znvol} that the equilibrium volume obtained from the total energy versus volume (E$-$V) curve for Zn(BH$_{4}$)$_{2}$(NH$_{3}$)$_{2}$ using GGA functional is (i.e., without vdW correction) overestimated by $-$3.33\% with respect to experimental value. However, the equilibrium volume obtained from the E$-$V curve based on the optPBE$-$vdW functional calculation deviates with the experimental values by only 0.31\% (see Table~\ref{tbl:1}). Chen \textit{et al.}~\cite{chen2018decomposition}considered van der Waals interaction using the optB86b$-$vdW functional and their calculated results show that the equilibrium volume deviate from experiment by 7.83\% and this is consistent with the present study. The TS$-$HI$-$vdW functional adequately describes the structure and the energetics for the ionic$-$like systems and hence we have also used this functional to predict the equilibrium structural parameters. Although the equilibrium volume obtained from TS$-$HI$-$vdW is underestimate by 3.18 \% over experiment, it predicts the equilibrium volume more reliably than GGA and optB8b$-$vdW. However, it is clear from Fig.~\ref{znvol} that the optPBE$-$vdW functional out perform to reliably predict structural parameters for transition metal ammine borohydrides such as Zn(BH$_{4}$)$_{2}$(NH$_{3}$)$_{2}$ than all other functionals considered in the present study. The current observation of accurate prediction of equilibrium structural parameter for molecular-like hydrides by optPBE$-$vdW is in consistent with our previous observation. \cite{kiruthika2019amphoteric}

\begin{table*}
\caption{Optimized equilibrium unit cell parameters (in \AA), equilibrium volume (in \AA $^{3}$) and the percentage of deviation of equilibrium volume from experiment ($\delta$ V in \%) for Zn(BH$_4$)$_2$(NH$_3$)$_2$ obtained from GGA and various van der Waals interaction included functionals. The lattice parameters are compared with the corresponding experimental~\cite{gu2012structure} and theoretical results.~\cite{chen2018decomposition}}
\label{tbl:1}
\begin{tabular*}{\textwidth}{@{\extracolsep{\fill}}cccccccc} 
\hline\\ [-1.5ex]
Particulars & & Experimental & GGA & optPBE$-$vdW & \multicolumn{2}{c}{optB86b$-$vdW} &  TS$-$HI$-$vdW\\
 \cline{6-7}
 & & & & & Present & Theory \cite{chen2018decomposition} & \\
\hline\\ [-1.5ex]
\textbf{Zn(BH$_4$)$_2$(NH$_3$)$_2$} \\[1.5ex]
Lattice Parameter (\AA)& a & 6.491 & 6.562 & 6.484 & 6.378 & 6.392 & 6.421\\[1.5ex]
& b & 8.887 & 8.984 & 8.877 & 8.731 & 8.417 & 8.791 \\[1.5ex]
& c & 6.462 & 6.532 & 6.455 & 6.349 & 6.388 & 6.392 \\[1.5ex]
& $\beta$ & 91.82 & 91.82 & 91.82 & 91.82 & 92.40 & 91.82 \\[1.5ex]
Cell Volume (\AA $^{3}$) & & 372.57 & 384.99 & 371.39 & 353.42 & 343.38 & 360.69\\[1.5ex]
$ \delta $V (\%)& & & $-$3.33 & 0.31 & 5.13 & 7.83 & 3.18 \\[1.5ex]
 \hline\\
\end{tabular*}
\end{table*}

\begin{table*}
\caption{The optimized atomic positions at the equilibrium volume obtained from GGA and optPBE$-$vdW for Mn(BH$_4$)$_2$(NH$_3$)$_2$ and Zn(BH$_4$)$_2$(NH$_3$)$_2$ are compared with the corresponding experimental values from \cite{gu2012structure} and \cite{jepsen2015tailoring}, respectively}
\label{tbl:2}
\begin{tabular*}{\textwidth}{@{\extracolsep{\fill}}lllll}
\hline
Compound & Atom and  & Experimental & GGA & optPBE$-$vdW\\ 
 and space group & Wyckoff position & & & \\ 
 & & & & \\
\hline\\
Mn(BH$_4$)$_2$(NH$_3$)$_2$;Pcab & Mn(8c) & 0.621, 0.149, 0.472 & 0.625, 0.131, 0.476  & 0.623,0.134, 0.972 \\
 & B1(8c) & 0.636, 0.077, 0.226 & 0.645, 0.063, 0.237  &  0.634, 0.098, 0.711\\
 & B2(8c) & 0.608, 0.969, 0.676 & 0.608, 0.956, 0.664  & 0.612, 0.956, 0.169 \\
 & N1(8c) & 0.529, 0.296, 0.482 & 0.535, 0.291, 0.483  & 0.532, 0.296, 0.984 \\
 & N2(8c) & 0.711, 0.291, 0.536 & 0.711, 0.280, 0.553  & 0.715, 0.281, 0.043 \\
 & H1(8c) & 0.532, 0.354, 0.577 & 0.530, 0.339, 0.587 & 0.527, 0.344, 0.091 \\
 & H2(8c) & 0.529, 0.359, 0.390 & 0.548, 0.371, 0.407 & 0.544, 0.379, 0.908 \\
 & H3(8c) & 0.482, 0.238, 0.483 & 0.482, 0.251, 0.453  & 0.477, 0.258, 0.956 \\
 & H4(8c) & 0.698, 0.338, 0.635 & 0.689,0.341, 0.641 & 0.698, 0.337, 0.139 \\
 & H5(8c) & 0.759, 0.235, 0.549 & 0.761, 0.235, 0.592 & 0.768, 0.232, 0.071\\
 & H6(8c) & 0.720, 0.366, 0.456 & 0.727, 0.352, 0.470 & 0.728, 0.357, 0.961 \\
 & H7(8c) & 0.633, 0.203, 0.242 & 0.635, 0.219, 0.148 & 0.624, 0.228, 0.691 \\
 & H8(8c) & 0.652, 0.050, 0.096 &  0.648, 0.015, 0.111 & 0.645, 0.034, 0.592 \\
 & H9(8c) & 0.575, 0.026, 0.256 & 0.578, 0.057, 0.301  & 0.282, 0.046, 0.776 \\
 & H10(8c) & 0.684, 0.029, 0.310 & 0.693, 0.072, 0.308 & 0.692, 0.077, 0.795 \\
 & H11(8c) & 0.671, 0.982, 0.626 & 0.674, 0.991, 0.613 & 0.676, 0.994, 0.113 \\
 & H12(8c) & 0.600, 0.044, 0.786 & 0.602, 0.995, 0.789 & 0.605, 0.012, 0.295 \\
 & H13(8c) & 0.597, 0.848, 0.712 & 0.610, 0.820, 0.651 & 0.611, 0.824, 0.170 \\
 & H14(8c) & 0.562, 0.003, 0.580 & 0.557, 0.996, 0.580 & 0.557, 0.997, 0.0861 \\
\hline
Zn(BH$_4$)$_2$(NH$_3$)$_2$;P12$ _{1} $1 & Zn(2a)& 0.122, 0.155, 0.880 & 0.124, 0.154, 0.880 & 0.133, 0.154, 0.877 \\
 & B1(2a) & 0.773, 0.158, 0.833 & 0.764, 0.161, 0.830 & 0.768, 0.161, 0.822\\
 & B2(2a) & 0.813, 0.642, 0.771 & 0.819, 0.652, 0.760 & 0.809, 0.652, 0.760\\                                     
 & N1(2a) & 0.699, 0.490, 0.255 & 0.704, 0.473, 0.252 & 0.700, 0.469, 0.256\\
 & N2(2a) & 0.764, 0.832, 0.293 & 0.744, 0.835, 0.291 & 0.739, 0.840, 0.291\\
 & H1(2a) & 0.738, 0.383, 0.211 & 0.749, 0.363, 0.205 & 0.758, 0.360, 0.218\\
 & H2(2a) & 0.547, 0.509, 0.212 & 0.546, 0.487,  0.212 & 0.543, 0.473, 0.209\\
 & H3(2a) & 0.712, 0.498, 0.414 & 0.717, 0.477, 0.416 & 0.706, 0.478, 0.419\\
 & H4(2a) & 0.888, 0.064, 0.755 & 0.876, 0.049, 0.776 & 0.880, 0.047, 0.771\\
 & H5(2a) & 0.864, 0.260, 0.929 & 0.865, 0.274, 0.902 & 0.872, 0.273, 0.895\\
 & H6(2a) & 0.679, 0.216, 0.690 & 0.673, 0.206, 0.668 & 0.679, 0.206, 0.660\\
 & H7(2a) & 0.670, 0.093, 0.956 & 0.650, 0.116, 0.970 & 0.654, 0.117, 0.962\\
 & H8(2a) & 0.828, 0.934, 0.257 & 0.791, 0.945, 0.244 & 0.779, 0.949, 0.234\\
 & H9(2a) & 0.606, 0.840, 0.273 & 0.581, 0.830, 0.281 & 0.575, 0.833, 0.289\\
 & H10(2a) & 0.796, 0.810, 0.447 & 0.787, 0.822, 0.449 & 0.790, 0.834, 0.447\\
 & H11(2a) & 0.890, 0.535, 0.868 & 0.889, 0.537, 0.859 & 0.879, 0.538, 0.862\\
 & H12(2a) & 0.657, 0.596, 0.687 & 0.658, 0.610, 0.666 & 0.647, 0.609, 0.669\\
 & H13(2a) & 0.759, 0.748, 0.880 & 0.766, 0.763, 0.873 & 0.758, 0.765, 0.873\\
 & H14(2a) & 0.943, 0.689, 0.658 & 0.962, 0.698, 0.649  & 0.952, 0.696, 0.649\\
\hline
\end{tabular*}
\end{table*}

\begin{table*}
  \caption{\ The equilibrium bond length (in \AA) between constituents in Zn(BH$_4$)$_2$(NH$_3$)$_2$ and Mn(BH$_4$)$_2$(NH$_3$)$_2$ obtained from GGA as well as optPBE$-$vdW functionals are compared with corresponding experimental values. The atom labels are same as those given in Table~\ref{tbl:1})}
  \label{tbl:3}
  \begin{tabular*}{\textwidth}{@{\extracolsep{\fill}}lllllllll}
    \hline\\
Mn(BH$_4$)$_2$(NH$_3$)$_2$ & & & & & Zn(BH$_4$)$_2$(NH$_3$)$_2$ & & & \\  
\cline{1-4} \cline{6-9}
Bond Interaction & \multicolumn{3}{c}{Bond Length (\AA)} & & Bond Interaction & \multicolumn{3}{c}{Bond Length (\AA)} \\
\cline{2-4} \cline{7-9}
& Experimental & GGA & optPBE$-$vdW & & & Experimental & GGA & optPBE$-$vdW\\ 
\hline\\
Mn$-$B1 & 2.310 & 2.313 & 2.314 & & Zn$-$B1 & 2.243 & 2.371 & 2.325 \\
Mn$-$B2 & 2.490 & 2.377 & 2.388 & & Zn$-$B2 & 2.255 & 2.371 & 2.323 \\                                     
Mn$-$N1 & 2.120 & 2.180 & 2.184 & & Zn$-$N1 & 2.014 & 2.158 & 1.034 \\
Mn$-$N2 & 2.149 & 2.176 & 2.181 & & Zn$-$N2 & 2.008 & 2.153 & 1.034  \\
H1$-$N1 & 1.011 & 1.034 & 1.034 & & H1$-$N1 & 0.978 & 1.067 & 1.034 \\
H2$-$N1 & 1.015 & 1.035 & 1.035 & & H2$-$N1 & 1.012 & 1.070 & 1.036   \\
H3$-$N1 & 1.000 & 1.036 & 1.035 & & H3$-$N1 & 1.016 & 1.069 & 1.036   \\
H4$-$N2 & 1.010 & 1.036 & 1.036 & & H4$-$B1 & 1.203 & 1.292 & 1.251  \\
H5$-$N2 & 1.002 & 1.037 & 1.036 & & H5$-$B1 & 1.184 & 1.284 & 1.246  \\
H6$-$N2 & 1.011  & 1.033 & 1.034 & & H6$-$B1 & 1.177 & 1.261 & 1.224  \\
H7$-$B1 & 1.201 & 1.227 & 1.231 & & H7$-$B1 & 1.184 & 1.263 & 1.225 \\
H8$-$B1 & 1.214 & 1.215 & 1.215 & & H8$-$N2 & 0.982 & 1.067 & 1.034  \\
H9$-$B1 & 1.200 & 1.252 & 1.252 & & H9$-$N2 & 1.011 & 1.070 & 1.036 \\
H10$-$B1 & 1.211 & 1.256 & 1.257 & & H10$-$N2 & 1.015 & 1.070 & 1.036 \\
H11$-$B2 & 1.194 & 1.253 & 1.253 & & H11$-$B2 & 1.182 & 1.286 & 1.248 \\
H12$-$B2 & 1.214 & 1.222 & 1.222 & & H12$-$B2 & 1.180 & 1.262 & 1.224 \\
H13$-$B2 & 1.203 & 1.223 & 1.224 & & H13$-$B2 & 1.195 & 1.290 & 1.250 \\
H14$-$B2 & 1.215 & 1.253 & 1.254 & & H14$-$B2 & 1.191 & 1.263 & 1.226 \\
\hline\\
  \end{tabular*}
\end{table*}

\subsubsection{Structural phase stability and ground state crystal structure}
In order to identify the ground state crystal structure of $M$(BH$_4$)$_2$(NH$_3$)$_2$ ($M$ = Sc, Ti, V, Cr, Fe, Co, Ni, and Cu), we have considered four potential structural variants such as Mg(BH$_4$)$_2$(NH$_3$)$_2$(Pcab), Ca(BH$_4$)$_2$(NH$_3$)$_2$(Pbcn), Sr(BH$_4$)$_2$(NH$_3$)$_2$ (Pnc2) and Zn(BH$_4$)$_2$(NH$_3$)$_2$ (P12$_{1}$1) mentioned in the above section(see section~\ref{stdsp}). Here, we substituted the above mentioned transition metals at the metal site and full geometry optimization was performed without any constraints on the atomic positions and cell parameters.

\begin{figure*}
 \centering
 \includegraphics[width=\linewidth]{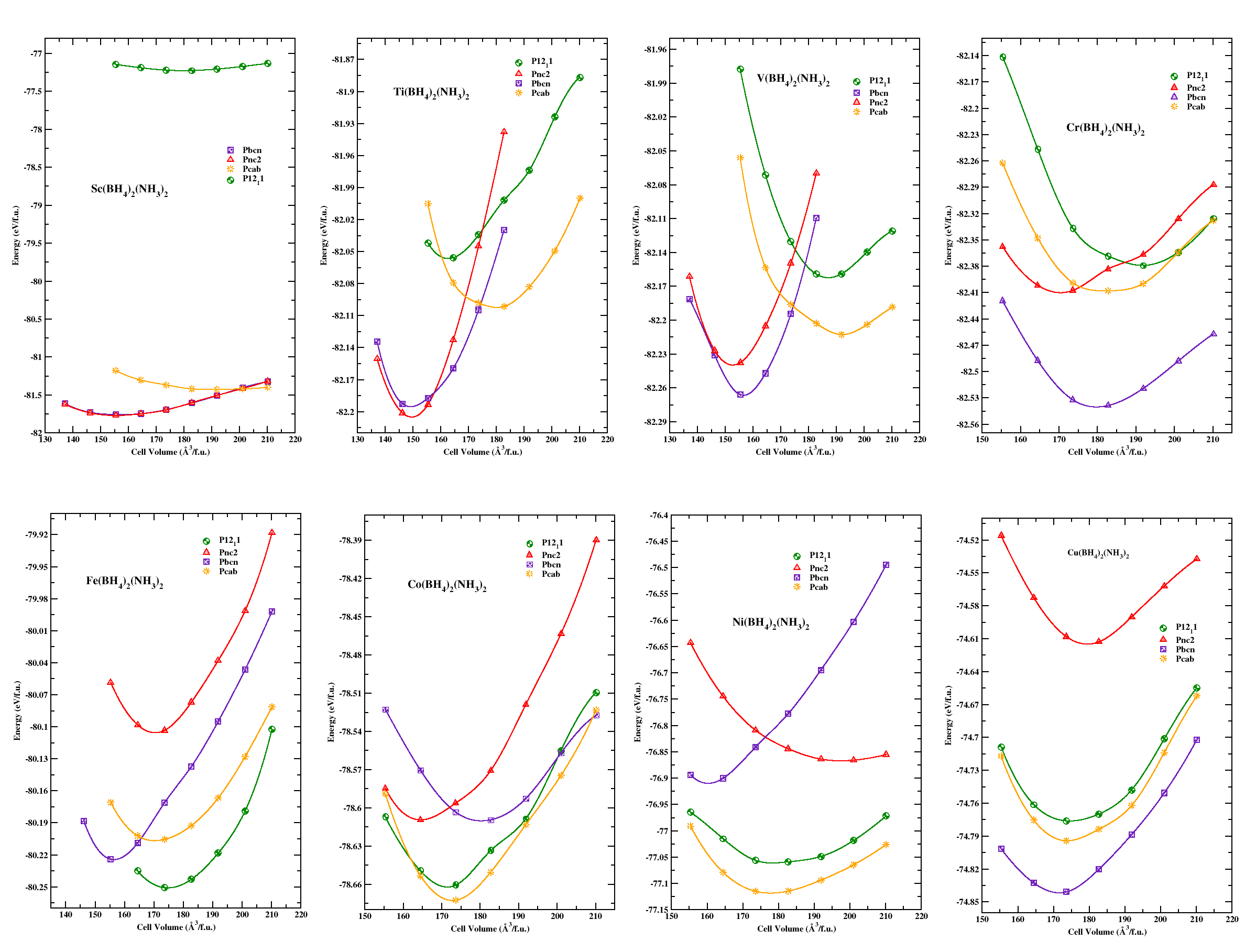}
 \caption{The calculated total energy vs. volume (E$-$V) curve for $M$(BH$_4$)$_2$(NH$_3$)$_2$ ($M$ = Sc, Ti, V, Cr, Fe, Co, Ni, and Cu) are obtained from \textit{ab$-$inito} calculation using optPBE$-$vdW functional. The E$-$V curve for all the structures are obtained using spin-polarised calculation considering ferromagnetic ordering.}
\label{EVFM}
\end{figure*}

The calculated total energy versus volume curve for $M$(BH$_4$)$_2$(NH$_3$)$_2$ ($M$ = Sc, Ti, V, Cr, Fe, Co, Ni, and Cu) are obtained from \textit{ab$-$inito} calculation using optPBE$-$vdW functionals. All these calculations were performed with spin-polarization, considering ferromagnetic ordering since the transition metal ions are magnetically polarised in these systems. Figure.~\ref{EVFM} shows the E$-$V curve for $M$(BH$_4$)$_2$(NH$_3$)$_2$ ($M$ = Sc, Ti, V, Cr, Fe, Co, Ni, and Cu). We have optimized the atomic coordinates and unit cell parameters globally using stress as well as force minimization with spin-polarised calculation considering ferromagnetic ordering in addition with the optPBE$-$vdW functionals, and thus identified the equilibrium cell volume, atomic coordinates, and unit cell dimensions which are listed in Table~\ref{tbl:5} along with the available experimental values. The equilibrium volume and bulk modulus were extracted from the calculated energy vs. cell volume curves by fitting them to the universal equation of state proposed by Vinet \textit{et al.}~\cite{vinet1986universal, vinet1989universal} and are virtually the same when obtained by fitting to the Brich ~\cite{birch1947finite} and Murnaghan~\cite{murnaghan1944compressibility} equation of states.

Here, Sc and Ti based TMABHs preferentially attain an orthorhombic structure with space group Pnc2 (see Fig.~\ref{EVFM}) as ground state. The TMABHs with transition metals, V, Cr, and Cu exhibit an orthorhombic structure with the space group Pbcn as the ground state. Among the ten TMABHs, the present study show that Fe/Zn(BH$_4$)$_2$(NH$_3$)$_2$ are exhibiting monoclinic structure with space group P12$_1$1 as evident from Fig.~\ref{EVFM} and ~\ref{cryst}. Along the 3\textit{d} transition metal based TMABHs series considered in the present study, Co and Ni based TMABHs preferentially attain an orthorhombic structure with space group Pcab as their ground state structure.

\begin{figure}
 \centering
 \includegraphics[width=\linewidth]{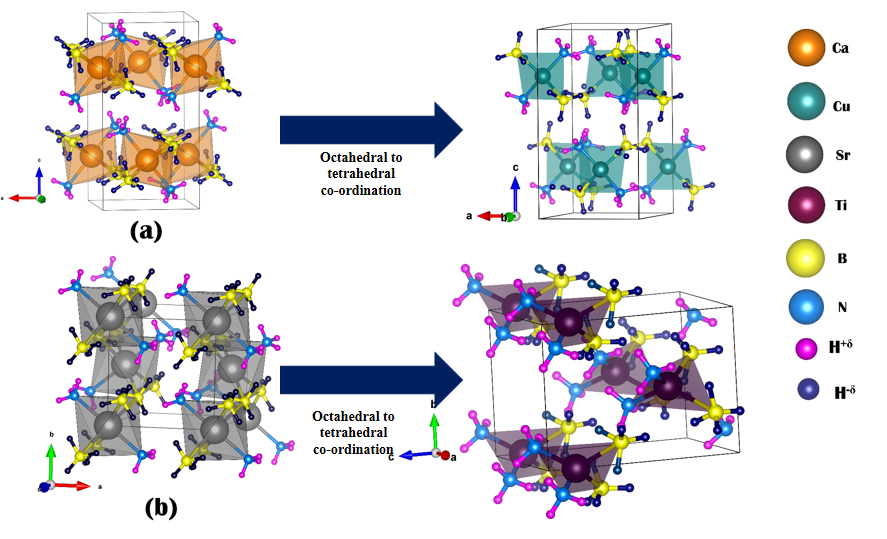}
 \caption{The orthorhombic Pbcn and Pnc2 starting structures obtained for alkaline$-$earth metal based AMABHs compounds transformed from octahedral to tetrahedral co$-$ordination when we replace alkaline$-$earth metal with transition metals as evident from the spin-polarised calculation using optPBE$-$vdW functional.}
\label{oct-tet}
\end{figure}

The optimized ground state structures obtained based on spin-polarised calculation suggest that the orthorhombic (Pcab) and monoclinic (P12$_1$1) phases have similar polyhedral co$-$ordination as we discussed in the structure description section(see section~\ref{stdsp}). On the other hand, the orthorhombic modifications with space group Pbcn ($M$ = V, Cr, Cu) and Pnc2 ($M$ = Sc, Ti) have distorted octahedral(i.e., four (BH$_4$)$^-$ groups bridges on the plane and two NH$_3$ groups axially over the plane) and tetrahedral (i.e., two (BH$_4$)$^-$ groups bridges on the plane and two NH$_3$ groups axially over the plane) co$-$ordinations, respectively and this deformation is due to a minimum variation in bond length between transition metal ion and molecular structural sub-units. In order to depict deformation of optimized ground state crystal structure of Pbcn ($M$ = V, Cr, Cu) and Pnc2  ($M$ = Sc, Ti) phases we have shown their crystal structures in Fig.~\ref{oct-tet}. Also, the bond length between transition metal ions and molecular structural sub-units and that between amphoteric hydrogen for $M$(BH$_4$)$_2$(NH$_3$)$_2$ ($M$ = Sc, Ti, V, Cr, Fe, Co, Ni, Cu) are tabulated in the Table~\ref{tbl:4}. When we compare the bond distance of AMABHs and TMABHs, as listed Table~\ref{tbl:4} its evident that when we replace alkaline$-$earth metals with transitions metals, the bond distance between the metal ion $M$ and BH$_4$ sub-units get reduced around 0.2~\AA ~to 0.5~\AA. Similarly, the bond distance between $M$ and NH$_3$ sub-unit also get reduced to around 0.3~\AA ~to 0.5~\AA. From the above observation it clear that, the  BH$_4$ and NH$_3$ structural sub-units are tightly bounded in TMABHs than that in AMABHs and hence one can expect that the release of borane and ammonia will happen at later stage than hydrogen release in TMABHs.

\begin{table}
\caption{The optimized equilibrium bond length (in \AA) between the transition metal ion with BH$_4$ as well as NH$_3$ molecular structural sub-units and that between amphoteric hydrides for $M$(BH$_4$)$_2$(NH$_3$)$_2$ ($M$ = Sc, Ti, V, Cr, Fe, Co, Ni, Cu) obtained using optPBE$-$vdW functional. For $A$(BH$_4$)$_2$(NH$_3$)$_2$ ($A$ = Mg, Ca, and Sr) systems the above mentioned bond lengths are compared with available corresponding experimental bond lengths in this Table.}
\label{tbl:4}
\begin{tabular}{llll} 
\hline\\ [-1.5ex]
Compound & \multicolumn{3}{c}{Bond Length (\AA)} \\ [1.5ex]
\cline{2-4} 
& Metal ($M$) $-$ NH$_3$ & Metal($M$) $-$ BH$_4$ & H$^{-\delta}$ $-$ H$^{+\delta}$ \\
\hline\\ [-1.5ex]
\textbf{Pcab}\\
\hline\\
Mg(BH$_4$)$_2$(NH$_3$)$_2$ & 2.156  & 2.127  & 1.040  \\[1.5ex]
& (2.150)\cite{soloveichik2008ammine} & (2.454) & (2.190) \\[1.5ex]
Co(BH$_4$)$_2$(NH$_3$)$_2$ & 2.082 & 2.218 & 1.893 \\[1.5ex]
Ni(BH$_4$)$_2$(NH$_3$)$_2$ & 2.061 & 2.165 & 1.848 \\[1.5ex]
\hline\\ [-1.5ex]
\textbf{Pbcn}\\
\hline\\
Ca(BH$_4$)$_2$(NH$_3$)$_2$ & 2.542  & 2.756  & 2.001  \\[1.5ex]
& (2.530)\cite{jepsen2015ammine} & (2.752) &  (2.001) \\[1.5ex]
V(BH$_4$)$_2$(NH$_3$)$_2$  & 2.141 & 2.281& 1.827 \\[1.5ex]
Cr(BH$_4$)$_2$(NH$_3$)$_2$ & 2.105 & 2.366 & 1.809 \\[1.5ex]
Cu(BH$_4$)$_2$(NH$_3$)$_2$ & 1.965 & 2.213 & 1.777 \\[1.5ex]
\hline\\ [-1.5ex]
\textbf{Pnc2}\\
\hline\\
Sr(BH$_4$)$_2$(NH$_3$)$_2$ & 2.703  & 2.868  & 1.952  \\[1.5ex]
 & (2.708)\cite{jepsen2015ammine} & (2.856) & (2.220) \\[1.5ex]
Ti(BH$_4$)$_2$(NH$_3$)$_2$ & 2.084 & 2.272 & 1.838\\[1.5ex]
Sc(BH$_4$)$_2$(NH$_3$)$_2$ & 2.216 & 2.486 & 1.946 \\[1.5ex]
\hline\\ [-1.5ex]
\textbf{P12$_1$1}\\
\hline\\
Fe(BH$_4$)$_2$(NH$_3$)$_2$ & 2.150 & 2.256 & 2.128\\[1.5ex]
\hline\\
\end{tabular}
\end{table}

\begin{table}
\tiny
\caption{The optimized equilibrium lattice parameters, equilibrium volumes and the magnetic moment($\mu_B$) at the transition metal sites, bulk modulus (B$_0$) and its pressure derivative (B$'_0$) for $M$(BH$_4$)$_2$(NH$_3$)$_2$ ($M$ = Sc, Ti, V, Cr, Mn, Fe, Co, Ni, and Cu) obtained from the ground state crystal structure using optPBE$-$vdW functional. The lattice parameters mentioned in brackets are from the corresponding experimental measurements reported in~\cite{jepsen2015ammine}.}
\label{tbl:5}
\begin{tabular*}{\textwidth}{@{\extracolsep{\fill}}lllllllll}
\hline
Compound & Space group & \multicolumn{3}{c}{Lattice Parameter(\AA)} & Equilibrium volume & Magnetic moment ($\mu_B$)& Bulk modulus(B$_0$) & B$'_0$ \\ 
\cline{3-5}
 & & a & b & c & (\AA $^{3}$/f.u.) & (with the magnetic ordering) & (GPa) & \\ 
\hline
Sc(BH$_4$)$_2$(NH$_3$)$_2$ & Pnc2 & 6.160 & 12.184 & 16.671 & 156.42 & 0.720  (G$-$AFM) & 16.00 & 8.04\\
Ti(BH$_4$)$_2$(NH$_3$)$_2$ & Pnc2 & 6.039 & 11.916 & 16.614 & 149.17 & 1.688  (A$-$AFM) & 16.66 & 5.53 \\
V(BH$_4$)$_2$(NH$_3$)$_2$ & Pbcn & 6.202 & 8.240 & 12.241 & 156.42 & 2.754  (A$-$AFM) & 10.85 & 1.67 \\
Cr(BH$_4$)$_2$(NH$_3$)$_2$ & Pbcn & 6.718 & 8.849 & 12.404 & 184.38 & 3.617  (C$-$AFM) & 7.45 & 5.65 \\
Mn(BH$_4$)$_2$(NH$_3$)$_2$ & Pcab & 17.483 (17.484) & 9.455 (9.455) & 8.872 (8.873) & 183.34 (183.36) & 4.338 (C$-$AFM) & 3.63 & 5.04\\
Fe(BH$_4$)$_2$(NH$_3$)$_2$ & P12$_{1}$1 & 12.289 & 16.534 & 6.153 & 156.23& 3.414 (C$-$AFM) & 4.03 & 7.56\\
Co(BH$_4$)$_2$(NH$_3$)$_2$ & Pcab & 17.040 & 9.395 & 8.703 & 174.19 & 2.415 (A$-$AFM) & 10.66 & 10.36 \\
Ni(BH$_4$)$_2$(NH$_3$)$_2$ & Pcab & 17.365 & 9.476 & 8.883 & 182.73 & 1.378 (C$-$AFM) & 7.60 & 7.01\\
Cu(BH$_4$)$_2$(NH$_3$)$_2$ & Pbcn & 6.431 & 8.755 & 11.810 & 166.29 & 0.502 (A$-$AFM) & 7.57 & 6.38\\
\hline
\end{tabular*}
\end{table}

In order to determine the phase stability and possible synthesis routes of TMABHs, we have calculated the reaction enthalpy of formation($\Delta$H$_{f}$). The reaction enthalpy of formation of compound is defined as the difference between the sum of the enthalpies of formation of the products $\Delta$H$_p$ (product) and the sum of the enthalpies of formation of the reactants $\Delta$H$_r$ (reactants). Considering the above statement the equation for $\Delta$H$_{f}$ is as follows:

\begin{equation}
\label{G1}
\Delta H_f = \sum \Delta H_p (product) - \sum \Delta H_r (reactants)
\end{equation}

If the sum of the enthalpy of formation ($\Delta$H$_p$) of product is less than the enthalpy of formation ($\Delta$H$_r$) of reactants, the enthalpy of reaction is negative. This negative value of $\Delta$H$_f$ would suggest that stable compound will form during the reaction and dissociation of such compound requires finite external thermal energy. In addition to this, the negative value of $\Delta$H$_f$ implies that the reaction is exothermic. If the enthalpy of reaction has the positive value of $\Delta$H$_f$ then those systems are unstable and the reaction is endothermic.

In the present study, we have considered four possible reaction pathways (Eq.~\ref{H1}$-$~\ref{H4}) and calculated the enthalpy of formation which are tabulated in Table~\ref{tbl:6}. The four reaction paths considered in the present study to calculate the enthalpy of formation ($\Delta$H$_{f}$) are as follows: 

\begin{equation}
\label{H1}
M + N_2 + 2B + 7H_2 \rightarrow M(BH_4)_2 (NH_3)_2
\end{equation}

\begin{equation}
\label{H2}
M + H_{2} + 2NH_{3} + B_{2}H_{6} \rightarrow M (BH_4)_2 (NH_3)_2
\end{equation}

\begin{equation}
\label{H3}
M + H_{2} + 2(NH_{3} \cdot BH_{3}) \rightarrow M (BH_4)_2 (NH_3)_2
\end{equation}

Using the calculated total energies obtained based on optimized ground state structural parameters we have estimated the heat of formation for stable phases of AMABHs and TMABHs (see Table~\ref{tbl:6}). The total energies of $M$(BH$_4$)$_2$(NH$_3$)$_2$ ($M$ = Mg, Ca, Sr, Sc, Ti, V, Cr, Mn, Fe, Co, Ni, Cu, and Zn), $M$, B$_{2}$H$_{6}$, NH$_{3}$, and H$_2$ are obtained from the geometry optimization calculation. The total energies for hydrogen and nitrogen molecules were computed via dimer in a larger cubic super$-$cell.

\begin{table*}
\caption{The calculated reaction enthalpy of formation($\Delta$H; in KJ mol$^{-1}$)  according to the reaction pathways (Eq. (\ref{H1})$-$(\ref{H3}) for the $M$(BH$_4$)$_2$(NH$_3$)$_2$ ($M$ = Mg, Ca, Sr, Sc, Ti, V, Cr, Mn, Fe, Co, Ni, Cu, and Zn) series obtained from optPBE$-$vdW functional.}
\label{tbl:6}
\begin{tabular}{l l l l} 
\hline\hline\\ [-1.5ex]
Compound & $\Delta$H$ _{1} $ & $\Delta$H$ _{2} $ & $\Delta$H$ _{3}$ \\
\hline\\ [-1.5ex]
Mg(BH$_4$)$_2$(NH$_3$)$_2$ & $-$335.41 & $-$85.21 &     106.38 \\[1.5ex]
Ca(BH$_4$)$_2$(NH$_3$)$_2$ & $-$528.49 & $-$278.29 &    131.44  \\[1.5ex]
Sr(BH$_4$)$_2$(NH$_3$)$_2$ & $-$758.46 & $-$508.26 & $-$262.66 \\[1.5ex]
\hline
Sc(BH$_4$)$_2$(NH$_3$)$_2$ & $-$488.23 &    238.01 & $-$7.58 \\[1.5ex]
Ti(BH$_4$)$_2$(NH$_3$)$_2$ & $-$363.51 &    113.29 & $-$32.30 \\[1.5ex]
V(BH$_4$)$_2$(NH$_3$)$_2$ & $-$313.34  &    63.12 & $-$182.47 \\[1.5ex]
Cr(BH$_4$)$_2$(NH$_3$)$_2$ & $-$291.82 &    41.02 & $-$204.58 \\[1.5ex]
Mn(BH$_4$)$_2$(NH$_3$)$_2$ & $-$247.23 & $-$497.43 & $-$742.94 \\[1.5ex]
Fe(BH$_4$)$_2$(NH$_3$)$_2$ & $-$219.82 & $-$30.39 & $-$276.00 \\[1.5ex]
Co(BH$_4$)$_2$(NH$_3$)$_2$ & $-$338.09 & $-$51.72 & $-$297.32 \\[1.5ex]
Ni(BH$_4$)$_2$(NH$_3$)$_2$ & $-$218.17 & $-$32.04 & $-$277.64  \\[1.5ex]
Cu(BH$_4$)$_2$(NH$_3$)$_2$ & $-$177.77 & $-$72.44 & $-$318.05 \\[1.5ex]
Zn(BH$_4$)$_2$(NH$_3$)$_2$ & $-$320.34 &    70.12 & $-$175.48 \\[1.5ex]
\hline \hline\\
\end{tabular}
\end{table*}

The reaction enthalpy of formation for Mg(BH$_4$)$_2$(NH$_3$)$_2$ and Ca(BH$_4$)$_2$(NH$_3$)$_2$ are exothermic for $\Delta$H$_1$ (Eq.~\ref{H1}) and $\Delta$H$_2$ (Eq.~\ref{H2}) reaction pathways. In the case of Sr(BH$_4$)$_2$(NH$_3$)$_2$, all the three considered reaction pathways are exothermic. From the overall observation we can state that almost all the reaction pathways are feasible for synthesising AMABHs except for reaction pathway (Eq.~\ref{H3})in Mg and Ca based AMABHs. The reaction enthalpy of Eq.~\ref{H1} can be considered as a heat of formation for the hydrides. So, if we compare the heat of formation for all the compounds considered in the present study we can conclude that these AMABHs compound can easily form from the reaction path in Eq.~\ref{H1} and also their highly stable.

In TMABHs the reaction enthalpy of $\Delta$H$_1$ (Eq.~\ref{H1}) and $\Delta$H$_3$ (Eq.~\ref{H3}) for all the TMABHs have high negative value. Similarly, the $\Delta$H$_2$ (Eq.~\ref{H2}) for Mn, Fe, Co, Ni, and Cu based TMABHs also have negative value. The observation of a negative value of $\Delta$H$_1$ indicates that these compounds are possible to synthesis with corresponding  reaction paths. The reaction enthalpy of $\Delta$H$_2$ (Eq.~\ref{H2}) for Sc, Ti, V, Cr and Zn based TMABHs have positive value indicating that this reaction pathway is energetically not possible to synthesis these compounds. Among the TMABHs series Mn, Fe, Co, Ni, and Cu based TMABHs systems can be synthesised using all the three reaction paths since all these three reactions give negative value of reaction enthalpy.

Comparing AMABHs and TMABHs, we observed that the reaction enthalpy ($\Delta$H$_1$) values of TMABHs are comparatively lower than that of AMABHs. This is due to the difference in the electronegativity value of alkaline$-$earth and transition metal ions in the systems. Electronegativity is a measure of ability of an atom to attract the electrons when the atom is part of a compound. The transition metal present in TMABHs system has low tendency to donate electrons to its neighbours, and this may leads to weak interaction between hydrogen and its neighbours i.e. B or N. Hence, such a systems can be dissociated easily and deliver the hydrogen in relatively low temperatures. The more detailed about the changes in the chemical bonding by varying cation with different electronegativity is discussed in section (~\ref{cba}).

\subsection{Magnetic Properties} \label{Mag}
A new class of magnetic compounds, molecular magnets~\cite{gatteschi1994large} has been attracting much attention towards the design of materials for novel technologies. For example most magnets are made of iron, cobalt, nickel or alloys of these materials. In order to make new magnetic materials with superior properties one has to exploit the flexibility of carbon chemistry that has been successful in producing the rich variety of biological systems found in nature; this approach leads to molecular magnets, i.e, magnetic materials in which the fundamental building block is the molecular unit rather than atomic unit.~\cite{gatteschi2006molecular, blundell2004organic, kahn1993molecular, miller2001magnetism} Each molecule of a compound is a nanomagnetic entity with a large spin or, in the antiferromagnet case, large staggered magnetization. The interaction between different molecules, being of the dipole$-$dipole type, is very small, so that the corresponding crystal is an arrangement of identical weakly interacting nanomagnets. An alternative strategy is to use transition metal or rare earth ions as the magnetic centre. But, non \textit{d/f}$-$block elements based groups between these centres to mediate the magnetic interactions. This results in enormous number of possible molecular architectures and combinations of ions. Molecular magnets are ideal objects to study phenomena of great scientific importance for mesoscopic physics, such as spin relaxation in nanomagnets, quantum tunneling of magnetization, topological quantum phase interference, quantum coherence, etc.

Olivier Kahn \cite{kahn1993molecular, postnikov2006density} was the first person to think about the effect of magnetism in molecular crystals. In particular, molecular magnetism deals with magnetic properties of isolated molecules and/or assemblies of molecules. Usually, the magnetic molecular crystal has one or more transition metal centers or rare$-$earth ions or simple organic radicals bound to their lattice sites by a stoichiometric chemical formula. The above statement agrees well with our TMABHs systems; that is, the isolated transition metals are centred with the two non$-$transition metal based molecular sub-units namely BH$_{4}$ and NH$_{3}$. Hence, TMABHs can be named as magnetic molecular crystals.

Magnetic properties of complex hydrides are very seldom studied experimentally and the magnetic properties of rare earth borohydrides were explored experimentally and theoretically in view of their magnetic cooling properties at cryogenic temperatures.~\cite{schouwink2016structural} A series of bimetallic Gd$-$borohydrides was investigated with the aim of identifying paramagnetic salts and investigating their properties related to the magnetic entropy change $\Delta$S$_M$, which is one of the key properties of magnetic refrigerants.~\cite{daudin1982thermodynamic} Moreover, a paradigm shift is currently taking place, attempting to develop Mn$^{2+}$ based materials for sub-Kelvin cooling.~\cite{tishin2016magnetocaloric} It may be noted that, the thermal conductivities of some simple metal borohydrides lie in the range of 2$-$5 Wm$^{-1}$ K$^{-1}$, which is appropriate for the heat transfer in magnetic cooling devices. \cite{sundqvist2006low, sundqvist2009thermal} So some the magnetic TMABHs systems considered in the present study may find application as magnetic refrigerants in ultra low temperatures. So, we have explored the magnetic properties of TMABHs in the present study.

\begin{figure}
 \centering
 \includegraphics[width=\linewidth]{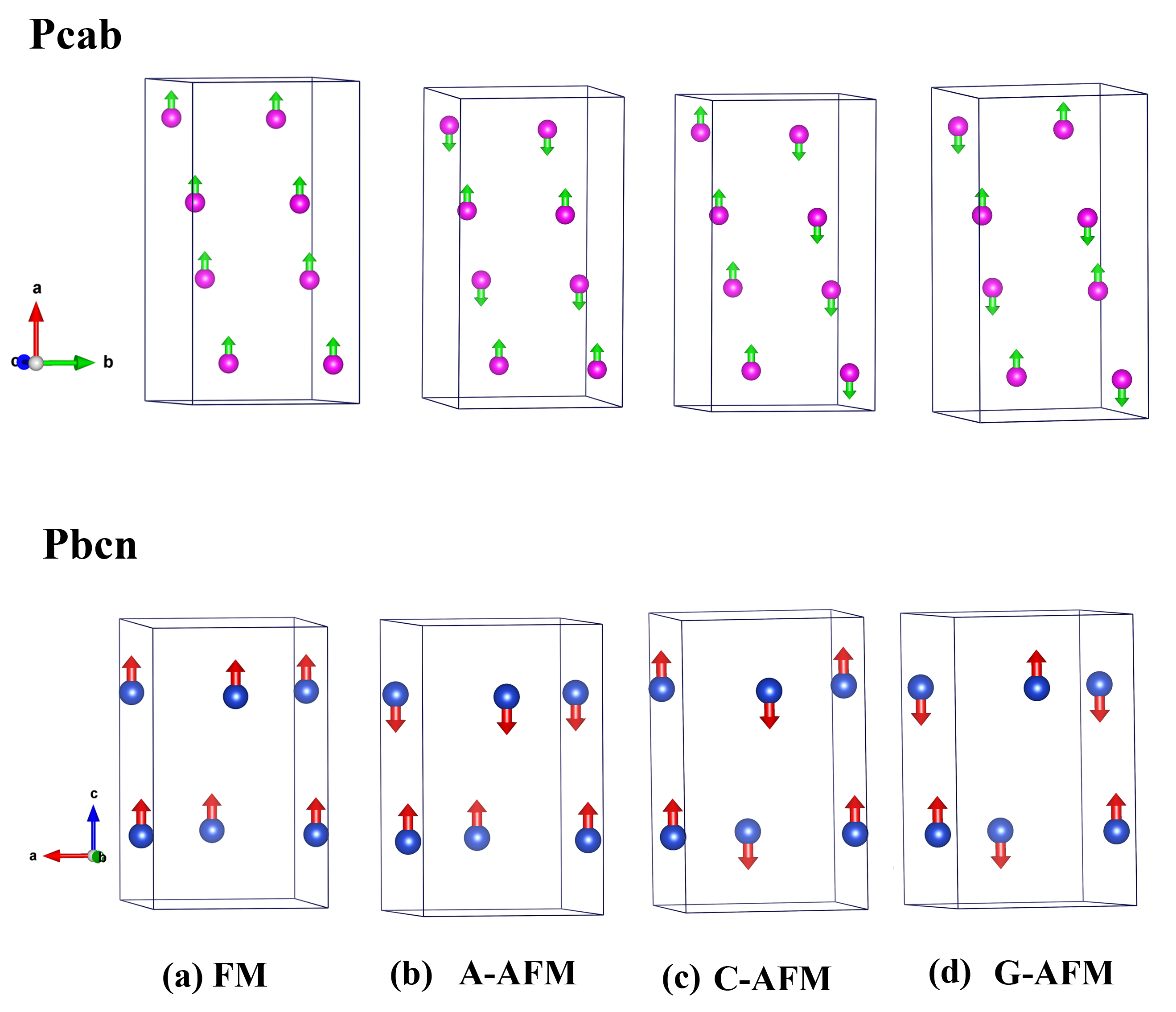}
 \caption{The schematic representation of (a) ferromagnetic (FM), and various antiferromagnetic (AFM) ordering such as (b) A$-$AFM, (c) C$-$AFM, and (d) G$-$AFM ordering in orthorhombic structure with space group Pcab (top row) and Pbcn (bottom row). Only transition metal atoms are shown for simplicity.}
 \label{AFMr}
\end{figure}

Using the optimized ground state crystal structural parameters, we have calculated the magnetic properties of TMABHs for the first time. To enunciate the ground state magnetic ordering for $M$(BH$_4$)$_2$(NH$_3$)$_2$ ($M$ = Sc, Ti, V, Cr, Mn, Fe, Co, Ni, and Cu) we have considered ferromagnetic (FM) and three types of possible antiferromagnetic (AFM) ordering (i.e., A$-$AFM, C$-$AFM and G$-$AFM) which are schematically shown in Fig.~\ref{AFMr}. In detail, A$-$AFM ordering occurs when the inter$-$plane coupling is antiferromagnetic and the intra$-$plane coupling is ferromagnetic. The C$-$AFM ordering are vice verse of A$-$AFM ordering i.e., the inter$-$plane coupling is ferromagnetic and intra$-$plane coupling is antiferromagnetic. Consistently, the G$-$AFM ordering both inter and intra$-$plane couplings are antiferromagtic. In order to account all the three possible AFM ordering we have created super$-$cell (2$\times$1$\times$2) for system having crystal structure with space groups Pnc2 and P12$_1$1.

\begin{figure*}
 \centering
 \includegraphics[width=\linewidth]{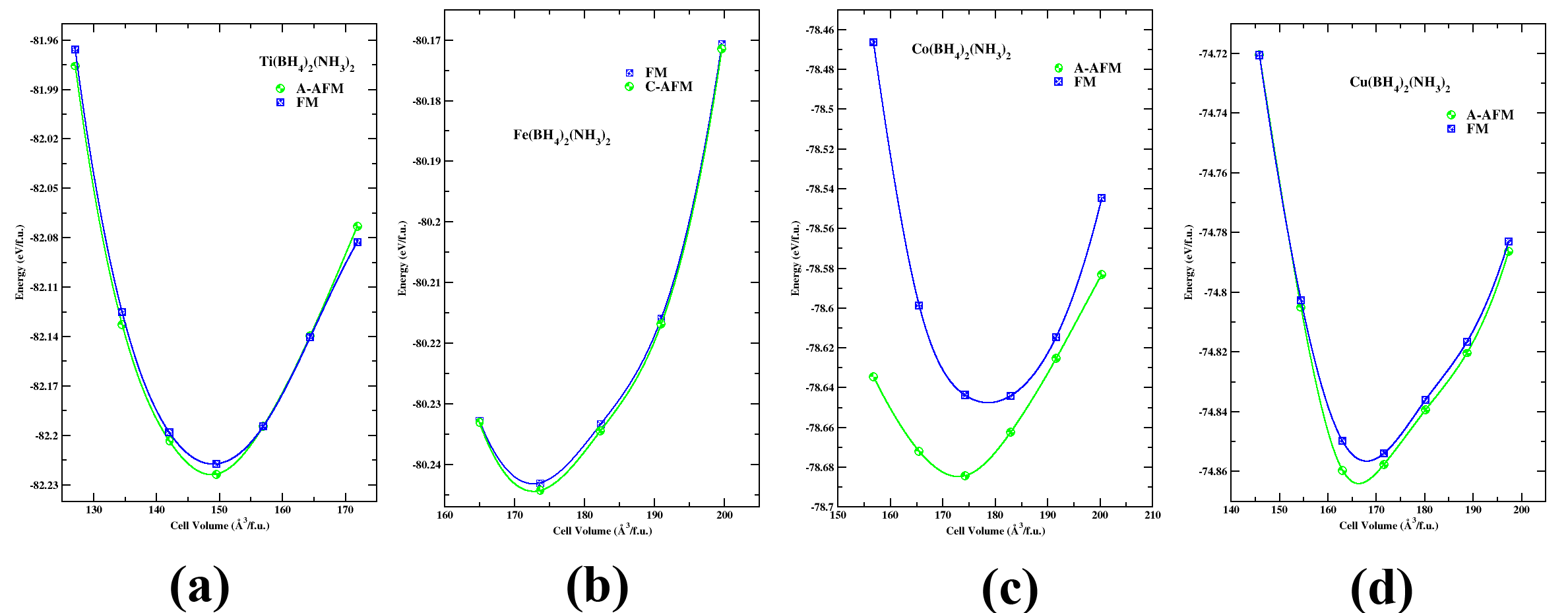}
 \caption{The calculated total energy vs. volume curve for ferromagnetic and ground state anti$-$ferromagnatic ordering for $M$(BH$_4$)$_2$(NH$_3$)$_2$ with (a) M = Ti (Pnc2), (b) M = Fe (P12$_1$1), (c) M = Co (Pcab), (d) M = Cu (Pbcn) ($M$ = Ti, Mn, Fe, Co, and Cu) as a representative systems.}
 \label{magvol}
\end{figure*}

From the calculated total energy for the TMABHs systems with the above mentioned various magnetic configurations we found that the antiferromagnetic ordering are stable configurations for all these systems and the total energy values with respect to magnetic ground state are listed in Table~\ref{tbl:7}. From the Table~\ref{tbl:7} we can confirm that Sc based TMABHs systems has G$-$AFM ordering as a ground state magnetic ordering. However, Ti, V, and Co based TMABHs systems attain the A$-$AFM ordering, whereas Cr, Mn, and Ni based TMABHs systems posses C$-$AFM ordering as their magnetic ground state. Comparing the total energy difference between the various magnetic configurations considered in the present study we found that the energy difference are very small compared with that is transition metal oxides. Due to the isolated nature of the magnetic ions in these molecular$-$like magnets there is weak exchange coupling between the ions and this could explains the small energy difference between various magnetic configurations. We have plotted the total energy versus volume curve for FM and ground state AFM configurations for $M$(BH$_4$)$_2$(NH$_3$)$_2$ ($M$ = Ti, Fe, Co, and Cu) as representative systems to substantiate that the TMABHs stabilize with AFM ordering as shown in Fig.~\ref{magvol} with small energy difference.

The Slater$–$Pauling rule states that adding an element to a metal compound will reduce the magnetization by a value proportional to the number of valence electrons~\cite{slater1936ferromagnetism, pauling1938nature, williams1983generalized, malozemoff1984band}. In order to substantiate the magnetic moment at the metal sites of TMABHs, we have plotted the magnetic moment ($\mu_{B}$) vs valence electron of 3\textit{d} transition metal in this series obtained from spin-polarised calculation which agrees well with the statement of Slater$-$Paulings rule. In addition to this, we have calculated the high spin magnetic moment based on the unpaired valence electrons (3\textit{d} transition metal ions) in the 2+ oxidation states using the following equation:

\begin{equation}
\label{MMUE}
\mu_{s} = \sqrt{n(n+1)} 
\end{equation}

\begin{figure}[h]
 \centering
 \includegraphics[scale= 0.8]{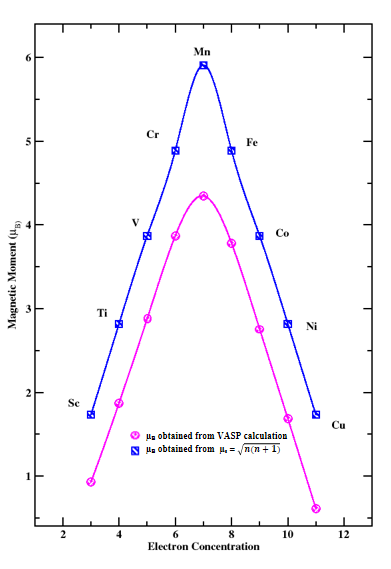}
 \caption{Variation in magnetic moment along the 3\textit{d} ions in TMABHs metal series which follow the Slater$-$Pauling curve behaviour. The pink color curve corresponds to SPC$-$1 and blue color curve corresponds to SPC$-$2}
 \label{SPC}
\end{figure}

Where $n$ is the valence electron of 3\textit{d} transition metal present in TMABHs. Slater$-$Paulings curve for TMABHs are shown in Fig.~\ref{SPC}. From the Fig.~\ref{SPC} it is evident that the magnetic moment vs. 3\textit{d} valence electrons obtained from the spin-polarized calculation form Slater$-$Paulings curve(SPC$-$1). Similarly, the calculated magnetic moment vs 3\textit{d} valence electrons obtained from the above equation also form the Slater$-$Paulings curve(SPC$-$2). The two Slater$-$Paulings curve (SPC$-$1 and SPC$-$2) are well matched with one another which enunciate the presence of magnetic ions in a high spin states. However, the calculated magnetic moments from our DFT calculations are always smaller than those obtained from Eq.~\ref{MMUE} indicating that some of these valence electrons participating in chemical bonding rather than magnetism though they are well isolated. The lattice parameters, equilibrium volume, magnetic moment and corresponding magnetic ordering for the ground state crystal structure of TMABHs are tabulated in Table~\ref{tbl:5}.  

Our total energy calculations predicted the magnetic ground state as A$-$AFM ordering (see Table~\ref{tbl:7}) for Ti(BH$_4$)$_2$(NH$_3$)$_2$. Usually, one can expect large magnetic moment in the FM configuration rather than AFM magnetic configuration. In the present study we found that the magnetic moment at the Ti site for the FM configuration is 1.720 $\mu_B$ whereas, in ground state A$-$AFM configuration magnetic moment is 1.688 $\mu_B$. The magnetic moment at the Ti site is influenced by the intra$-$planar FM coupling between the Ti atoms. It is expected that, the presence of localised magnetism due to weak interaction with the neighbours in this system, the inter$-$planar FM coupling will be weaker. It is to be noted that, the inter$-$planar AFM coupling is very important to understand the C$-$AFM in this system. The C$-$AFM and G$-$AFM configuration of Ti(BH$_4$)$_2$(NH$_3$)$_2$ has magnetic moment of 1.715 $\mu_B$ and 1.717 $\mu_B$, respectively at the Ti site. The calculated magnetic moment at the Ti site in various magnetic configurations are almost the same as evident from the Table~\ref{tbl:7}. It is beyond doubt that the valence electrons at the Ti is in high spin state all the magnetic configurations due to its present in the molecular like structural sub-unit.

\begin{figure}
\centering
 \includegraphics[scale= 0.2]{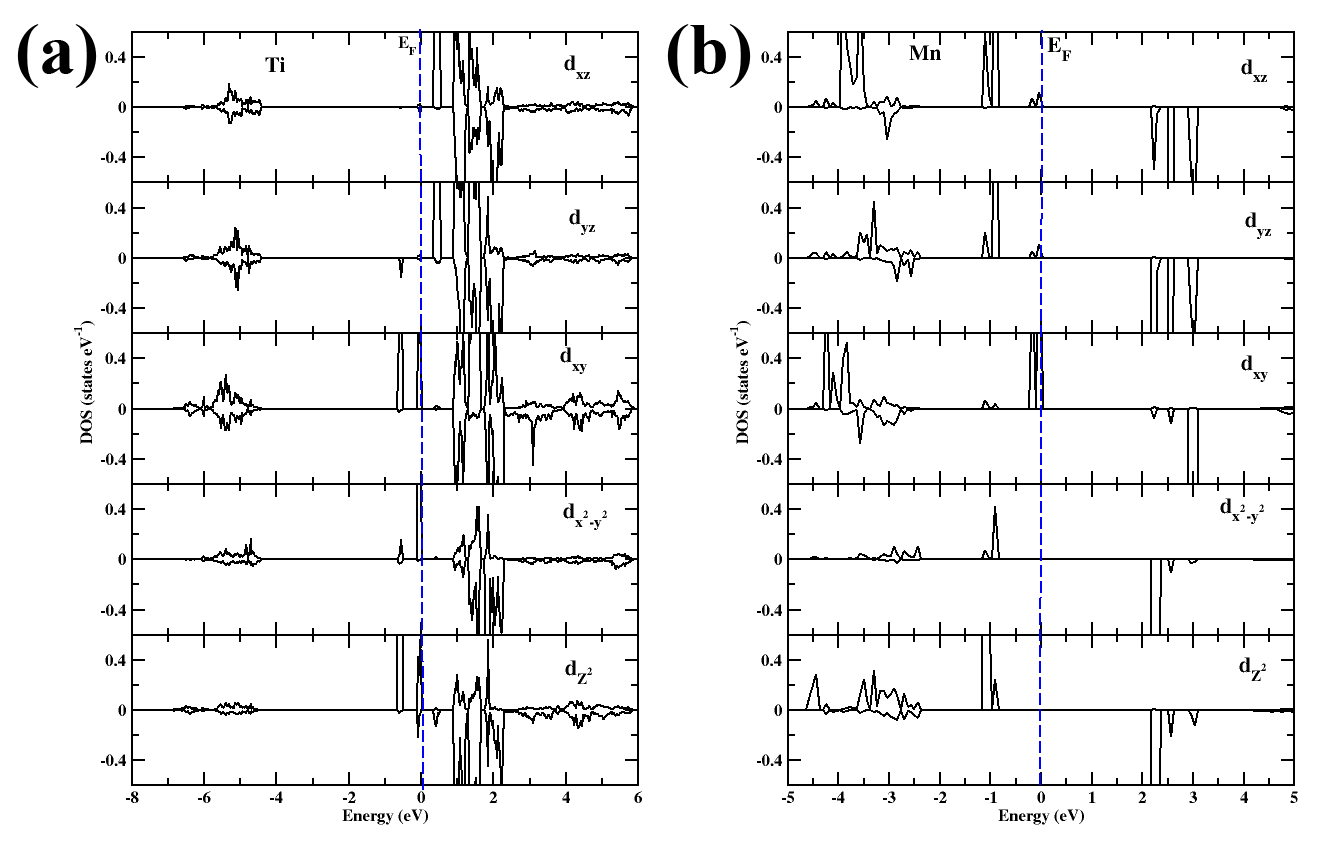}
 \caption{The spin-polarised orbital projected density of states for $M$(BH$_4$)$_2$(NH$_3$)$_2$ ($M$ = Ti, Mn) as representative systems obtained from optPBE$-$vdW functional at the equilibrium volume. The Fermi level is set to zero.}
 \label{odos}
\end{figure}

In order to substantiate the magnetic properties of TMABHs, we have plotted the spin-polarised orbital decomposed DOS for the 3\textit{d} states of transition metal ions in $M$(BH$_4$)$_2$(NH$_3$)$_2$ ($M$ = Ti, Mn, Fe, Co, and Cu) as representative systems obtained in their magnetic ground state and are shown in Fig.~\ref{podos}. In an ideal tetrahedral cubic crystal field, the \textit{d}$-$level splits into doubly degenerate \textit{e} (d$_{x^2-y^2}$, d$_{z^2}$) and triply degenerate \textit{t}$_2$ (d$_{xy}$, d$_{xz}$, d$_{yz}$) levels. If the system is purely ionic, Ti$^{2+}$ with 2 electrons in the valence shell will fill the energy levels with a low spin state (d$^2_{z^2}$, d$^0_{x^2-y^2}$, \textit{t}$^0_2$) and high spin state (d$^1_{z^2}$, d$^1_{x^2-y^2}$, \textit{t}$^0_2$) with spin moments 0 $\mu_B$ and 2 $\mu_B$ respectively. As it can be seen from our orbital projected DOS for ground state A$-$AFM configuration given in Fig.~\ref{odos}(a), it is clearly evident that Ti$^{2+}$ ions are occupied in the majority spin channel (d$_{z^2}$, d$_{x^2-y^2}$, and d$_{xy}$) with 2 electrons. However due to orthorhombic distortion the degeneracy of \textit{d} states are further lifted as evident from Fig.~\ref{odos} (a). We have calculated the occupation number of electrons in the majority spin channel of these orbitals and are d$_{z^2}$ = 0.84, d$_{x^2-y^2}$ = 0.70, d$_{xy}$ = 0.32, d$_{yz}$ = 0.10, d$_{xz}$ = 0.08 and the sum of electron occupation in these five orbitals is $\sim $ 2. Hence the pure ionic picture is well described the occupation number as well as the high spin magnetic moment discussed above.

Similarly, here we describe the magnetic properties of Mn(BH$_4$)$_2$(NH$_3$)$_2$. Our spin-polarised total energy calculations predict that C$-$AFM  as magnetic ground state (see Table~\ref{tbl:7}) for Mn(BH$_4$)$_2$(NH$_3$)$_2$. The Mn$^{2+}$ ion with the tetrahedrally co$-$ordinated with ligands in pure ionic picture will have 5 electrons fill the \textit{d} levels a low spin state (d$^2_{z^2}$, d$^2_{x^2-y^2}$, \textit{t}$^1_2$,) or high spin state (d$^1_{z^2}$, d$^1_{x^2-y^2}$, \textit{t}$^3_2$) with spin moments 1 $\mu_B$ and 5 $\mu_B$, respectively. In the present study we found that the magnetic moment at the Mn sites for FM configuration is 4.347 $\mu_B$. Whereas, in ground state C$-$AFM configuration magnetic moment at the Mn site is 4.342 $\mu_B$. As expected, the magnetic moment in Mn site for FM configuration is higher than the AFM configurations. Similar to M = Ti systems the Mn based TMABHs also having orthorhombic crystal structure with tetrahedrally co$-$ordinated transition metal ion. And hence \textit{d}$-$states will split similar to that in M = Ti case. However, due to the increasing the magnetic moment in Mn site bring C$-$AFM type magnetic ordering as ground state. But, it may be noted that the energy difference between A$-$AFM and C$-$AFM ordering is 1.7 meV/f.u. only due to very weak magnetic coupling within the plane and between the planes due to well isolated nature of transition metal ions. So, all these magnetic configurations the Mn ions are in the high spin state only and also the calculated difference in the magnetic moment at the Mn site in various magnetic configuration are very small (see Table~\ref{tbl:7}).

From the Fig.~\ref{odos}(b) we found that, all the 5 \textit{d} orbitals are  almost half filled indicating that Mn$^{2+}$ ion is in high spin state. Also, the well localised nature of \textit{d} states in the valence band indicating the presence of molecular-like behaviour in this system. Due to the formation of covalent bonding by the Mn ions with the neighbours some  of its bonding states are filled up in the down spin channel around $-$3 eV and hence instead of 5 $\mu_B$/Mn expected for the high spin state we are getting slightly lower value of 4.33 $\mu_B$/Mn for its C$-$AFM magnetic ground state.

\begin{table*}
\caption{The calculated total energy ($\Delta$E)(meV/f.u.) relative to the lowest energy states and their corresponding magnetic moment for $M$(BH$_4$)$_2$(NH$_3$)$_2$ ($M$ = Sc, Ti, V, Cr, Mn, Fe, Co, Ni, Cu) in the FM, A$-$AFM, C$-$AFM, G$-$AFM and non$-$magnetic (NM) configurations obtained from spin$-$polarised calculation. The relative energy are given with respective to that of the magnetic ground state. }
\label{tbl:7}
\begin{tabular*}{\textwidth}{@{\extracolsep{\fill}}ccccccccccc} 
\hline\\ [-1.5ex]
Compound Name & \multicolumn{5}{c}{$\Delta$E (meV/f.u.)} & & \multicolumn{4}{c}{Magnetic moment ($\mu_B$)}   \\
\cline{2-6} \cline{8-11}
& FM & A$-$AFM & C$-$AFM & G$-$AFM & NM & & FM & A$-$AFM & C$-$AFM & G$-$AFM\\
\hline\\	
Sc(BH$_4$)$_2$(NH$_3$)$_2$ & 8.2 & 0.7 & 0.8 & 0 & 158 & & 0.786 & 0.746 & 0.719 & 0.720 \\[1.5ex]
Ti(BH$_4$)$_2$(NH$_3$)$_2$ & 6 & 0 & 0.7 & 0.2 & 372 & & 1.720 & 1.688 & 1.715 & 1.717\\[1.5ex]
V(BH$_4$)$_2$(NH$_3$)$_2$ & 4.6 & 0 & 5.9 & 10.3 & 1348 & & 2.783 & 2.754 & 2.743 & 2.744\\[1.5ex]
Cr(BH$_4$)$_2$(NH$_3$)$_2$ & 5.2 & 5.1 & 0 & 0.5 & 1871 & & 3.622 & 3.612 & 3.617 & 3.614 \\[1.5ex]
Mn(BH$_4$)$_2$(NH$_3$)$_2$ & 18.6 & 1.7 & 0 & 0.8 & 5849 & & 4.347 & 4.344 & 4.342 & 4.343\\[1.5ex]
Fe(BH$_4$)$_2$(NH$_3$)$_2$ & 1.3 & 1.5 & 0 & 0.2 & 457 & & 3.417 & 3.409 & 3.414 & 3.412\\[1.5ex]
Co(BH$_4$)$_2$(NH$_3$)$_2$ & 41.1 & 0  & 1.7 & 2.4 & 750 & & 2.404 & 2.412 & 2.379 & 2.378\\[1.5ex]
Ni(BH$_4$)$_2$(NH$_3$)$_2$ & 7.1 & 3.2 & 0 & 1.5 & 614 & & 1.373 & 1.381 & 1.378 & 1.380 \\[1.5ex]
Cu(BH$_4$)$_2$(NH$_3$)$_2$ & 7.8 & 0 & 4.6 & 0.1 & 128.2 & & 0.509 & 0.502 & 0.508 & 0.500 \\[1.5ex]
\hline\\
\end{tabular*}
\end{table*}

\subsection{Chemical Bonding Analysis} \label{cba}
In order to understand the chemical bonding interaction between the constituents in $M$(BH$_4$)$_2$(NH$_3$)$_2$ ($M$ = Sc, Ti, V, Cr, Mn, Fe, Co, Ni, Cu, Zn) the density of states (DOS), charge density distribution, electron localization function (ELF), Born effective charge analyses (BEC) and Bader effective charge (BC) analyses were made. We will have brief discussion about the above set analyses in the subsequent sections.
\subsubsection{Density of States}
\begin{figure}[h]
 \centering
 \includegraphics[scale= 0.6]{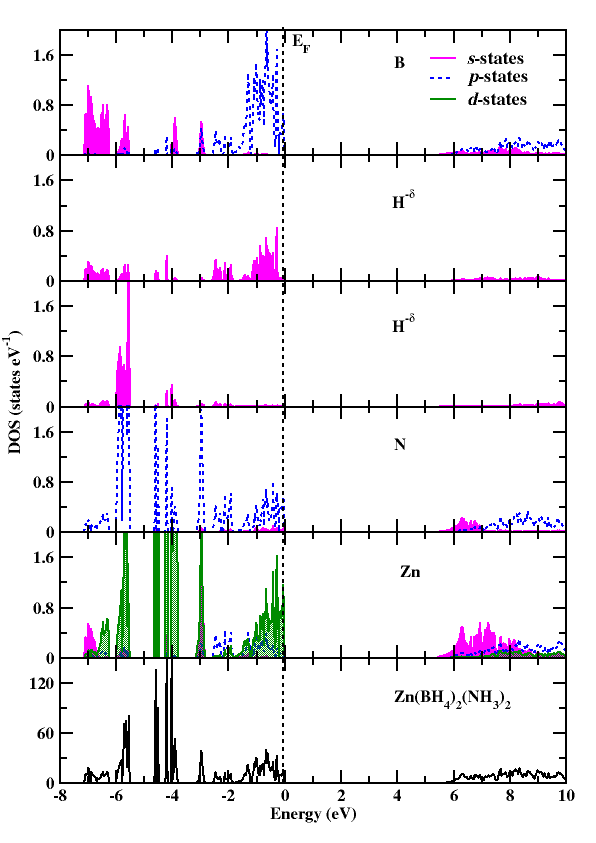}
 \caption{The total and partial density of states for Zn(BH$_4$)$_2$(NH$_3$)$_2$ obtained from optPBE$-$vdW functional at the equilibrium volume. The Fermi level is set to zero.}
 \label{zndos}
\end{figure}

The total density of states (TDOS) and partial DOS (PDOS) for Zn(BH$_4$)$_2$(NH$_3$)$_2$ are shown in Fig.~\ref{zndos}. From Fig.~\ref{zndos} we can confirm that the Zn(BH$_4$)$_2$(NH$_3$)$_2$ is an insulator with a band gap of 5.3 eV. It may be noted that the estimated band gap based on GGA calculation usually underestimate the experimental value by 30\% to 50 \%  and hence one could expect the experimental band gap will be much higher than this value. Considering the partial DOS distribution in the valence band, the Zn$-$\textit{d} and N$-$\textit{p} states are present in the entire valence band. Whereas, top of the valence band (VB) is equally contributed by Zn$-$\textit{d} and B$-$\textit{p} states with noticeable contribution from both N$-$\textit{p} and H$^{-\delta}$ $-$ \textit{s} states. The PDOS distribution of Zn and N show that the Zn$-$\textit{d} and N$-$\textit{p} states are energetically degenerated in entire VB, which indicates the presence of a covalent interaction between Zn and N. Similarly, the DOS distribution of Zn$-$\textit{d} and B$-$\textit{p} states are energetically degenerated from $-$2.0 eV to VB maxima and this distribution reveals the presence of covalent interaction between the Zn and B. Examining the DOS distribution of N and H$^{+\delta}$, due to high electronegative nature of N it draw charges from neighbouring hydrogen site and hence negligible charges are present at the H site higher energy range of the VB. This implies the presence of strong ionic bonding between them. In contrast, the DOS distribution of B and H$^{-\delta}$ shows that the B$-$\textit{p} and H$^{-\delta}$ $-$ \textit{s} states are energetically degenerate from $-$2 eV to the VB maxima indicating the presence of a substantial covalent bond between them. Thus, we conclude that the covalent interaction between B and H$^{-\delta}$ is stronger than that between N and H$^{+\delta}$.

It is interesting to note from Fig.~\ref{zndos} we can see that the DOS at the H closer to B is completely different form that closer to N and this clearly indicating that the hydrogen is in two different oxidation states in this materials. Comparing the PDOS distribution between H$^{-\delta}$ and H$^{+\delta}$, it is clearly evident that \textit{s}$-$states of H$^{+\delta}$ are well localised in the narrow energy range from $-$6 eV to $-$5.5 eV. If we integrate the PDOS of H$^{+\delta}$ in the entire valence band then we found that the number of electrons in the H$^{+\delta}$ site is much smaller than that of the neutral H atom indicating that this hydrogen is present in a positive oxidation state. On the other hand, H$^{-\delta}$ $-$\textit{s} states are present throughout the VB with a strong peak around the vicinity of $-$2 eV. Moreover, the presence of a relatively higher amount of electrons in H$^{-\delta}$ states than in those of the neutral H atom indicates that this hydrogen is in a negative oxidation state. From these observation we can conclude the presence of amphoteric hydrogen in this system which is further established by various charge analyses discussed later. 

\begin{figure}[h]
 \centering
 \includegraphics[scale= 0.6]{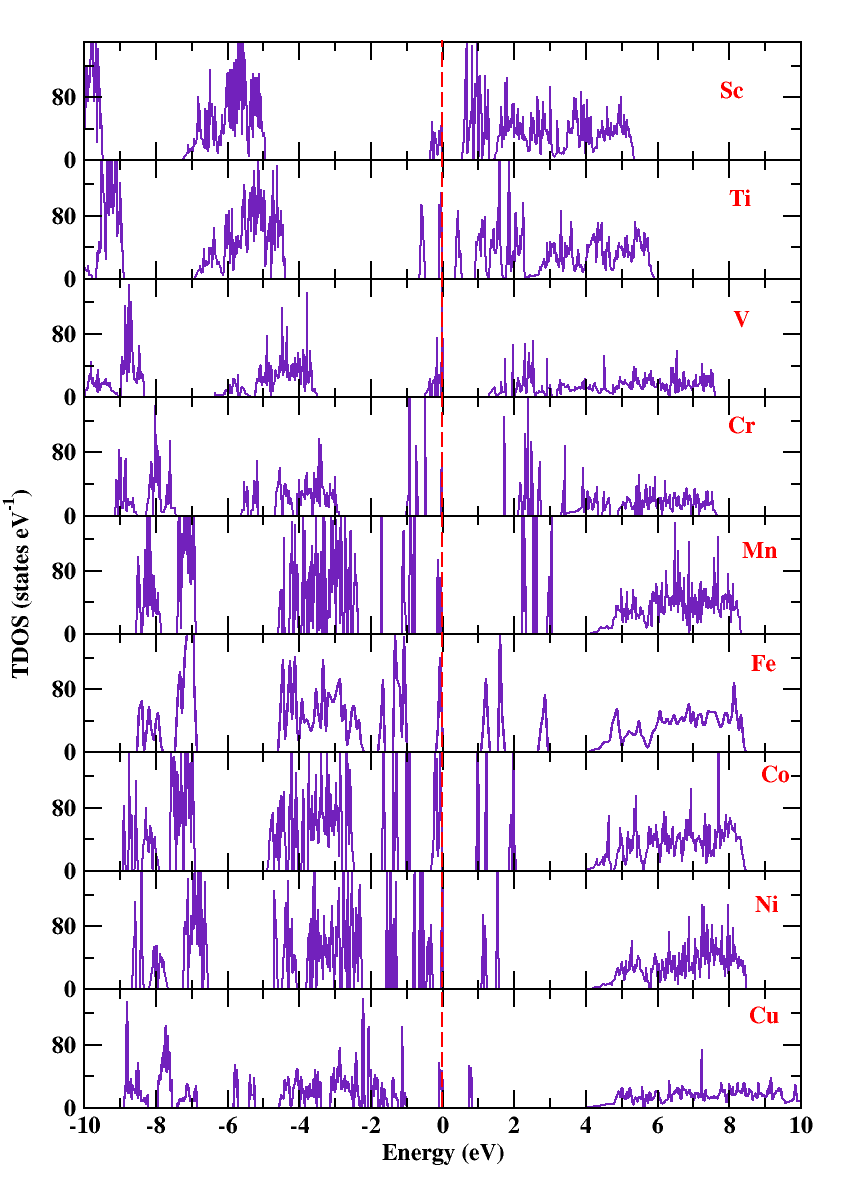}
 \caption{The total density of states (TDOS) for $M$(BH$_4$)$_2$(NH$_3$)$_2$ ($M$= Sc, Ti, V, Cr, Mn, Fe, Co, Ni, and Cu) obtained from optPBE$-$vdW functional at the equilibrium volume in their magnetic ground state. The Fermi level is set to zero.}
 \label{tdostm}
\end{figure}

The calculated TDOS at the equilibrium volume for the ground$-$state structures with lowest energy antiferromagnetic configuration of $M$(BH$_4$)$_2$(NH$_3$)$_2$ ($M$ = Sc, Ti, V, Cr, Mn, Fe, Co, Ni, Cu) compounds are shown in Fig.~\ref{tdostm}. From Fig.~\ref{tdostm}, it is clearly evident that all these compounds have finite energy gap between the valance band maxima and conduction band minima indicating that all are antiferromagnetic insulators. The calculated band gap values from the TDOS for $M$(BH$_4$)$_2$(NH$_3$)$_2$ with $M$ = Sc, Ti, V, Cr, Mn, Fe, Co, Ni, and Cu are 0.50, 0.34, 1.27, 1.69, 2.19, 1.02, 0.91, 1.45, and 0.72 eV, respectively. We found that the vicinity of band edges (above $-$2 eV and below $-$4 eV) the electron distribution is almost molecular-like in all these systems. Due to formation of molecular-like complex in these systems.  Considering the TDOS in the VB region, due to increasing the number of \textit{d} electron when go from Sc to Cu the Fermi level is systematically shifted towards higher energy as evident in this Fig.~\ref{tdostm}. Compared with transition metal alloys the \textit{d}states are well localised that brings magnetism with high spin states in these systems.

Considering partial DOS distribution of constituents from Ti(BH$_4$)$_2$(NH$_3$)$_2$, the valence band is mainly originated from the Ti$-$\textit{d}, N$-$\textit{p}, and B$-$\textit{p} states with minor contribution form the Ti$-$\textit{s}, \textit{p}, N$-$\textit{s}, B$-$\textit{s} and H$^{-\delta}$ $-$ \textit{s} states. owing to the isolated nature of Ti cation in this system it \textit{d} states are well localised that brings strong exchange splitting and hence high spin state. Due to the presence of strong ionic bonding between the constituents, the high energy region of the VB (above $-$4.0 eV) is almost empty. Also, the transfer of electron from H$^{+\delta}$ ions makes eligible small DOS distribution in the H$^{+\delta}$ site.

\begin{figure}
\centering
 \includegraphics[scale= 0.17]{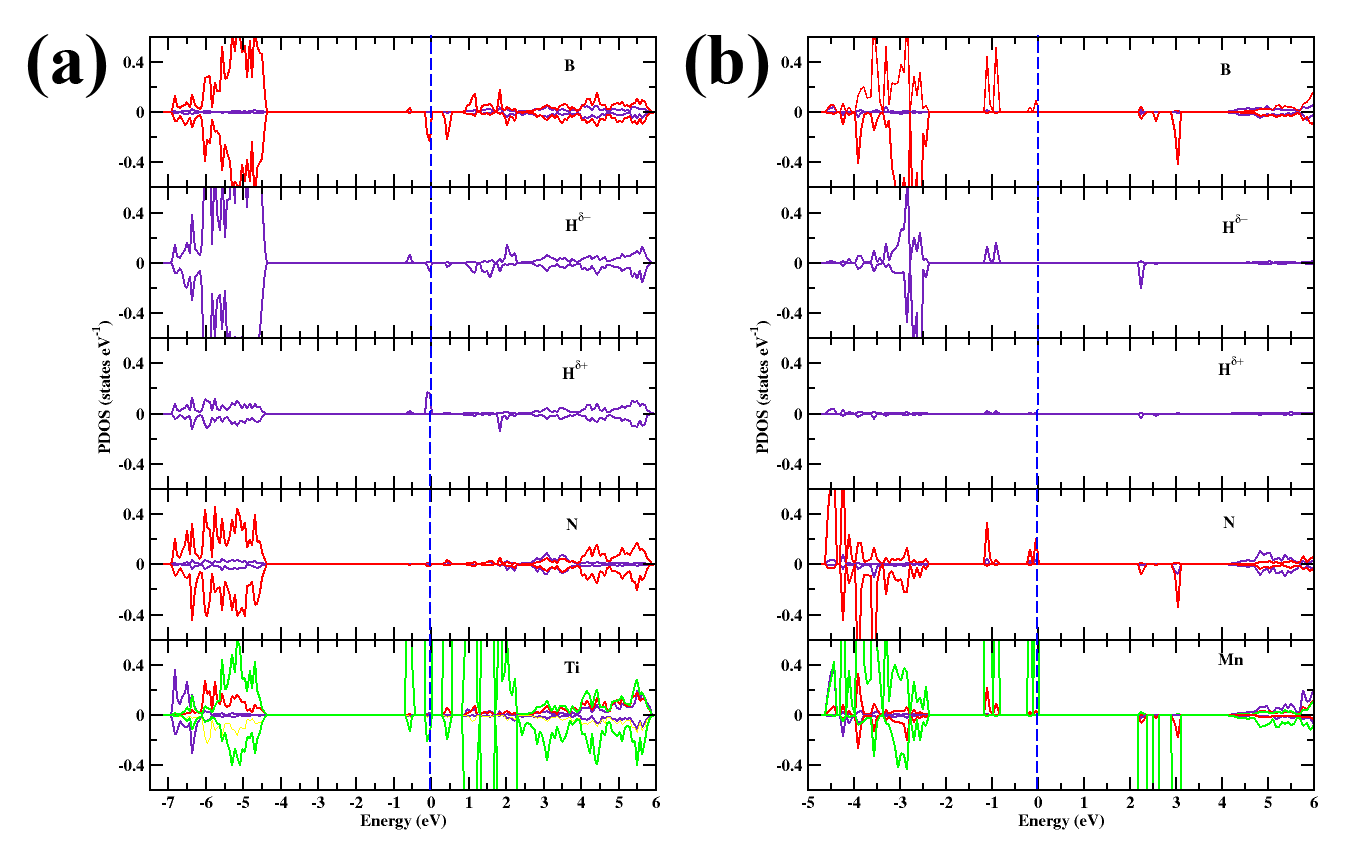}
 \caption{The partial density of states (PDOS) for $M$(BH$_4$)$_2$(NH$_3$)$_2$ $M$ = Ti and Mn are shown (a) and (b), respectively obtained from optPBE$-$vdW functional at the equilibrium volume. The Fermi level is set to zero.}
 \label{podos}
\end{figure}

The PDOS distribution for Mn(BH$_4$)$_2$(NH$_3$)$_2$ shown in Fig.~\ref{podos} (b), the valence band is mainly originated from the Mn$-$\textit{d}, N$-$\textit{p}, B$-$\textit{p} states with minor contributions from the Mn$-$\textit{s, p}, N$-$\textit{s}, B$-$\textit{s} and H$^{-\delta}$ $-$\textit{s} states. Due to the presence of noticeable covalent bonding between the N and H$^{+\delta}$ as well as B and H$^{-\delta}$ there is substantial energetically degenerated DOS distribution present in the higher energy region of the VB ( above $-$2.5 eV) originating from these atoms.  Due to amphoteric behaviour of hydrogen ions we could see the distinct DOS distribution at the H$^{+\delta}$ and H$^{-\delta}$ sites. The Mn$^{2+}$ ions are in well isolated states with huge exchange splitting energy of more than 2 eV brings high spin states with magnetic moment of 4.34 $\mu_B$. Due to the presence of covalent bonding between Mn with N as well as B brings substantial DOS distribution at the minority spin channel of Mn sites in the energy range $-$2.5eV to $-$4.5 eV and this reduce the magnetic spin moment from 5 $\mu_B$ to 4.34 $\mu_B$. Further the covalent bonding interaction between B and H is much stronger than that between the N and H and hence the bonding states of B$-$\textit{p}$-$H$^{-\delta}$ $-$\textit{s} bonding hybrid is present in the entire VB as evident from the Fig.~\ref{podos} (b). From these PDOS analysis one can conclude that the TMABHs have both ionic as well as covalent interaction between the constituents. Hence, the similar kind of bonding behaviours can be observed for the remaining TMABHs systems where the DOS distribution are given in the supplementary information.

\subsubsection{Charge Density Analysis}
\begin{figure*}
 \includegraphics[scale= 0.07]{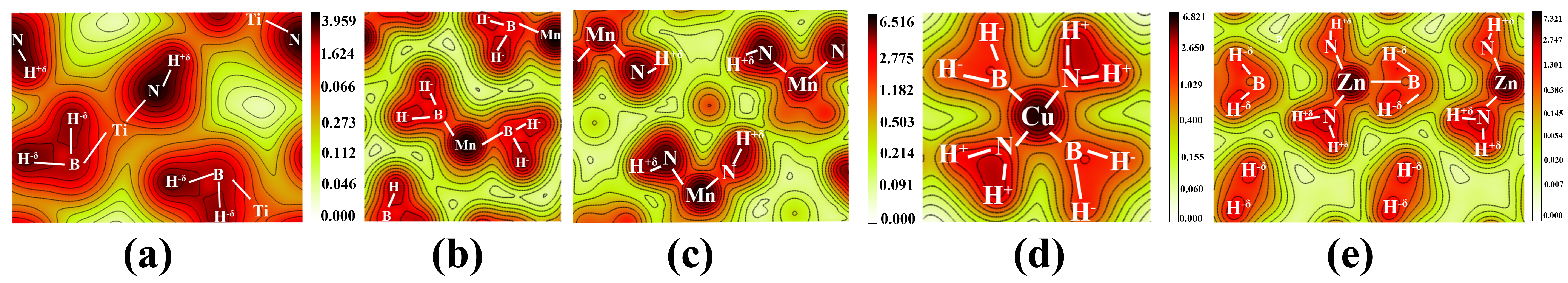}
 \caption{The calculated charge density distribution between constituents in $M$(BH$_4$)$_2$(NH$_3$)$_2$ with $M$ = Ti, Mn, Cu, and Zn are given in (a), (b and c), (d), and (e),  respectively. The charge density (unit of e/\AA $^3$) is projected in a plane where both B$-$H and N$-$H bonds are available.}
 \label{chg}
\end{figure*}

In order to substantiate the amphoteric behavior of hydrogen and also the presence of finite covalent bonding between amphoteric hydrogen ions with their neighbours in TMABHs. We have shown charge density distribution for $M$(BH$_4$)$_2$(NH$_3$)$_2$ ($M$ = Ti, Mn, Cu, and Zn) in Fig.~\ref{chg} as representative system for each structural modification.

We have shown charge density distribution of Zn(BH$_4$)$_2$(NH$_3$)$_2$ in Fig.~\ref{chg} (d). Usually, the spherical distribution of charges around the atomic sites and the negligibly small charge between the constituents describe the ionic bonding. On the other hand, the anisotropic charge distribution at the atomic sites and finite charge distribution between the constituents are the indication for covalent bonding. Between B and H$^{-\delta}$ as well as N and H$^{+\delta}$ there is a large charge distribution is present and also they are having anisotropic nature indicating substantial covalent bonding between H with neighbours. Moreover, the hydrogen ions are partially covalently bonded with them self those near the N sites as well as B sites. These feature could explain why only partial charge is present at the H sites. Also due to this strong covalent bonding, We can visualise this structure with molecular-like BH$_{4}$ and NH$_{3}$ structural sub-units. The charge at the Zn sites are spherical distributed with substantial amount of charge is polarised towards N as well as B sites indicating the bonding interaction between Zn with its neighbour is of iono$-$covalent nature.

However, the hydrogen that closer to the B has relatively more spherical distribution with relatively small charge distribution between B and H$^{-\delta}$ enunciate the noticeable ionic bonding character. Due to nearly protonic state of hydrogen associated with N the inter atomic distance between N and H$^{+\delta}$ (see Table~\ref{tbl:4}) is shorter than that between B and H$^{-\delta}$ and the charge is merely accumulated in the N site within the NH$_3$ structural sub-units. 

The calculated charge density distribution in the planes where one could see bonding interaction between constituents in Mn(BH$_4$)$_2$(NH$_3$)$_2$ are given in the Fig.~\ref{chg}(b and c). Though the charge density distribution at the Mn site is nearly spherically symmetric there is substantial anisotropic charge distribution pointing towards N as well as B indicating the presence of iono$-$covalent bond between Mn and the N/B. Let us analysis the bonding interaction within the BH$_4$ structural sub-units. There is finite charge distributed anisotropically around B and H$^{-\delta}$ site indicating the presence of finite covalent bonding between B and H$^{-\delta}$. Due to strong electronegativity nature of N it draws most of the charge from the neighbouring hydrogen making the H$^{+\delta}$ almost protonic like character as a consequence of this we are unable to see noticeable charge at the H$^{+\delta}$ site. Further, the bonding interaction between N and H$^{+\delta}$ is more ionic than that between any of the other constituents one can expect negligible charge between N and H $^{+\delta}$. However, due to the protonic nature of the H $^{+\delta}$ the charge density distribution plot show that the N and H$^{+\delta}$ are strongly attached to each other, a typical feature one could see the charge density plot of protonic hydrogen with its neighbours. In contrast, the hydrogen neighbouring to the B sites are having H$^{-\delta}$ oxidation state and substantial amount of charge is present in the H$^{-\delta}$ sites. Similar kind of charge density distribution are observed in the other TMABHs considered in the present study. So, one can conclude that the bonding interaction between N$-$H$^{+\delta}$ and B$-$H$^{-\delta}$ is covalent with noticeable ionic character. In order to quantify the covalent and the ionic character of bonding interaction between constituents in TMABHs we have consulted various charge analysis schemes discussed as follows.

\subsubsection{Electron localisation Function}
In order to enunciate the amphoteric behavior of hydrogen and also the presence of finite covalent bonding between hydrogen with its neighbours, we have plotted the electron localisation function (ELF) obtained from optPBE$-$vdW interaction included spin-polarised calculation. ELF analysis help in understanding the empirical concept of localised electrons, specially the pair electron localization in the spirit of Lewis structures~\cite{vajeeston2008structural}. Fig.~\ref{elf} shows the calculated ELF for M = Ti, Mn, Cu and Zn based TMABHs.

The smaller value of ELF at the transition metal sites $M$ ($M$ = Ti, Mn, Cu and Zn) clearly show that there is depletion of electron at the $M$ sites, indicating the presence of ionic interaction between the metal $M$ and the structural sub-units BH$_4$ and NH$_3$. There is substantial amount of ELF value present between B$-$H$^{-\delta}$ and N$-$H$^{+\delta}$ suggesting that there is noticeable covalent bonding interaction present between hydrogen with its neighbours. In order to have a clear picture of the amphoteric behaviour of hydrogen (H$^{-\delta}$ and H$^{+\delta}$) present in these systems, we have plotted the iso$-$surface for the ELF with the value of  0.95 for Cu based TMABHs as a representative system in Fig.~\ref{elf}. From this figure it is clear that the ELF distribution for hydrogen closer to N has a smaller distribution  than that closer to B. So, the three$-$dimensional visualization of the ELF iso$-$surface shows beyond doubt the presence of hydrogen with amphoteric behaviour in these systems.

\begin{figure*}
 \centering
 \includegraphics[scale= 0.07]{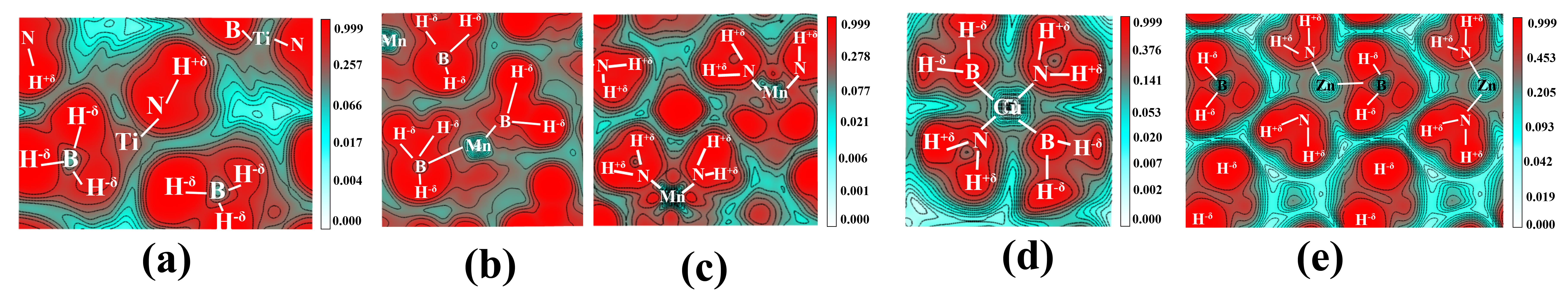}
 \caption{The calculated valance$-$electron ELF distribution for $M$(BH$_4$)$_2$(NH$_3$)$_2$ $M$ = Ti, Mn, Cu, and Zn are given in (a), (b and c), (d), and (e) respectively. The ELF is projected in a plane where both B$-$H and N$-$H bonds are clearly visible.}
 \label{elf}
\end{figure*}

\begin{figure}
 \centering
 \includegraphics[scale= 0.08]{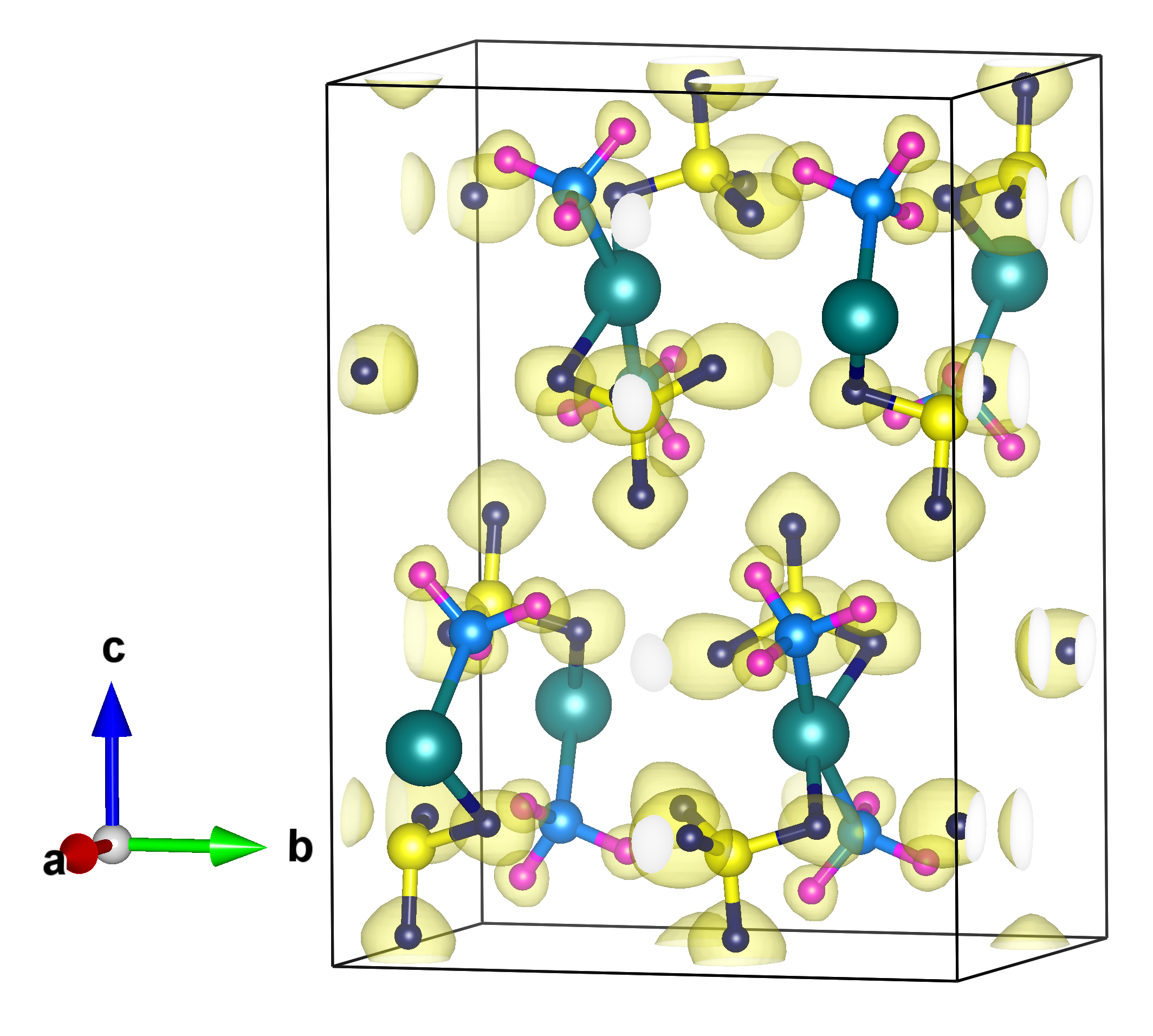}
 \caption{Isosurface (value of 0.95) of the valence$-$electron localization function for Cu(BH$_4$)$_2$(NH$_3$)$_2$ as a representative system for TMABHs where the hydrogen associated with N has smaller ELF distribution than that associated with B indicating the ampoteric behaviour of hydrogen.}
 \label{elf}
\end{figure}

\subsubsection{Born$-$Effective charge analysis}
In order to have deeper insight in to the bonding interaction between the constituents and the amphoteric behaviour of hydrogen, we have calculated the Born effective charge (BEC) at the various atomic sites and their values are listed in Table~\ref{BEC}. In order to calculate the Born effective charge tensor we have used the Berry phase approach of the "Modern Theory of Polarisation". ~\cite{king1993theory} Let us consider the diagonal components of BEC tensor of constituents in TMABHs system where the metal ion ($M$ = Ti, Mn, Fe, Cu, and Zn) has lower BEC value than the nominal charge of 2$+$ and also the diagonal components are not equal which implies the presence of covalent interaction between the metal atom with the neighbouring structural sub$-$units such as BH$_4$ and NH$_3$. However, off$-$diagonal components of BEC at the metal sites ($M$ = Ti, Mn, Fe, Cu, and Zn) are very small indicating  that the covalency effect is minimal. Hence, from the above observation it is clear that the bonding between $M$ and its neighbours has iono$-$covalent bonding in TMABHs systems.

The average diagonal components of BEC at B, N, H$^{-\delta}$ and H$^{+\delta}$ are 0.38, $-$0.70, $-$0.29 and 0.25, respectively and these values are smaller than the corresponding nominal ionic charge of $+$3, $-$3, $-$1 and $+$1, respectively. Further, the diagonal components of BEC tensor for  B, N, H$^{-\delta}$ and H$^{+\delta}$ also not same for all three direction suggesting the presence of anisotropic charge distributions and hence covalent bonding is also expected to present within the structural sub-units. The presence of finite value of off$-$diagonal components of BEC tensor also re$-$iterating presence of covalency. Interestingly, as expected from DOS, charge density, and structural analysis, the hydrogen closer to B has negative BEC value and that closer to N has the positive BEC value confirming the presence of amphoteric hydrogen in TMABHs systems.                                               

\begin{table*}
\small
\centering
\caption{The calculated Bader charge (BC) and Born effective charge tensor ($Z^*_{ij}$) for the constituents in $M$(BH$_4$)$_2$(NH$_3$)$_2$ ($M$ = Ti, Mn, Fe, Cu and Zn) obtained from the spin-polarised optPBE$-$vdW calculation.}
\label{BEC}
\begin{tabular}{c c* {9}{c c c c c c c c c}}
\hline\hline\\ [-1.5ex]
Compound & Atom site & BC & \multicolumn{9}{c}{$Z^*$(e)}\\
\cline{4-12} \\[-1.5ex]
 & & & Z$_{xx}$ & Z$_{yy}$ & Z$_{zz}$ & Z$_{xy}$ & Z$_{yz}$ & Z$_{zx}$ & Z$_{xz}$ & Z$_{zy}$ & Z$_{yx}$\\
 \hline\hline\\ [-1.5ex]
 Ti(BH$_4$)$_2$(NH$_3$)$_2$ & Ti & 1.371 & 1.600 & 2.483 & 1.325 & $-$0.017 & 0.009 & 0.000 & 0.001 & 0.000 & 0.602 \\[2.5ex]
  & B & 1.580 & 0.283 & 0.272 & 0.606 & $-$0.237 & 0.010 & 0.025 & 0.137 & 0.020 & $-$0.070\\[2.5ex]
 & H$^{-\delta}$ & $-$0.571 & $-$0.373 & $-$0.275 & $-$0.226 & 0.030 & $-$0.151 & 0.194 & 0.091 & $-$0.058  & 0.133  \\ [2.5ex]
 & N & $-$1.428 & $-$0.765 & $-$0.696 & $-$0.641 & 0.161 & $-$0.139 & 0.080 & $-$0.074 & $-$0.039 & 0.112\\ [2.5ex]
 & H$^{+\delta}$ & 0.523 & 0.224 & 0.238 & 0.310 & 0.052 & 0.050 & 0.036 & 0.026 & 0.080 & 0.104\\ [2.5ex]
\hline\\
Mn(BH$_4$)$_2$(NH$_3$)$_2$ & Mn & 1.254 & 1.426 & 1.736 & 1.884 & 0.048 & $-$0.039 & 0.035 & $-$0.004 & $-$0.032 & 0.030\\[2.5ex]
  & B & 1.505 & 0.098 & 0.171 & 0.072 & 0.008 & $-$0.038 & 0.072 & 0.008 & $-$0.115 & $-$0.019\\[2.5ex]
 & H$^{-\delta}$& $-$0.549 & $-$0.178 & $-$0.456 & $-$0.258 & 0.009 & 0.055 & 0.002 & 0.019 & 0.047 & 0.034 \\ [2.5ex]
 & N & $-$1.447 & $-$0.748 & $-$0.771 & $-$0.499 & 0.137 & $-$0.091 & 0.003 & $-$0.002 & $-$0.042 & 0.108\\ [2.5ex]
 & H$^{+\delta}$ & 0.544 & 0.301 & 0.195 & 0.320 & 0.012 & 0.069 & 0.089 & 0.035 & 0.016 & 0.059\\ [2.5ex]
\hline\\
 Fe(BH$_4$)$_2$(NH$_3$)$_2$ & Fe & 1.212 & 1.374 & 1.789 & 2.010 & $-$ 0.012& $-$0.044 & 0.086 & 0.121 & $-$0.016 & 0.024\\[2.5ex]
  & B & 1.548 & 0.171 & 0.230 & 0.305 & $-$0.062 & 0.048 & $-$0.032 & $-$0.003 & $-$0.016 & $-$0.059\\[2.5ex]
 & H$^{-\delta}$ & $-$0.572 & $-$0.220 & $-$0.257 & $-$0.432 & 0.081 & 0.101 & $-$0.136 & $-$0.106 & 0.130 & 0.072\\ [2.5ex]
 & N & $-$1.504 & $-$ 0.738 & $-$0.795 & $-$0.596 & 0.151  & 0.042 & $-$0.020 & $-$0.070 & 0.075 & 0.074\\ [2.5ex]
 & H$^{+\delta}$ & 0.543 & 0.286 & 0.186 & 0.326 & $-$0.021 & 0.058 & 0.095 & 0.053 & $-$0.027 & $-$0.002\\ [2.5ex]
\hline\\
Cu(BH$_4$)$_2$(NH$_3$)$_2$ & Cu & 0.866 & 1.269 & 0.440 & 1.314 & 0.000 & 0.000  & 0.036 & 0.053 & 0.000 & 0.000 \\[2.5ex]
  & B & 1.535 & 0.452 & $-$0.023 & 0.515  & $-$0.224 & $-$0.020 & $-$0.139 & $-$0.053 & $-$0.155 & $-$0.045 \\[2.5ex]
 & H$^{-\delta}$ & $-$0.538 & $-$0.708 & $-$0.067 & $-$0.340 & $-$0.047 & 0.016 & $-$0.025 & $-$0.007 & 0.057 & 0.003 \\ [2.5ex]
 & N & $-$1.351 & $-$0.455 & $-$0.337 & $-$0.460 & $-$0.045 & 0.029 & 0.196 & 0.203 & $-$0.015 & $-$0.062 \\ [2.5ex]
 & H$^{+\delta}$ & 0.482 & 0.153 & 0.399 & 0.322 & $-$0.048 & 0.171 & 0.050 & 0.020 & 0.342 & $-$0.002 \\ [2.5ex]
\hline\\
Zn(BH$_4$)$_2$(NH$_3$)$_2$ & Zn & 1.030 & 1.967 & 1.598 & 1.957 & 0.054 & 0.057 & $-$0.050 & $-$0.046 & 0.048 & 0.054\\[2.5ex]
  & B & 1.645 &0.200 & 0.092 & 0.127 & $-$0.010 & $-$0.082 & 0.016 & $-$0.003 & 0.000 & $-$0.026\\[2.5ex]
 & H$^{-\delta}$ & $-$0.552 & $-$0.391 & $-$0.284 & $-$0.114 & 0.236 & $-$0.098 & 0.010 & 0.046 & $-$0.063 & 0.122\\ [2.5ex]
 & N & $-$1.664 & $-$0.826 & $-$0.763 & $-$0.804 & 0.097 & $-$0.092 & 0.125 & $-$0.001 & $-$0.066 & 0.113
\\ [2.5ex]
 & H$^{+\delta}$ & 0.571 & 0.301 & 0.233 & 0.294 & $-$0.009 & 0.075 & $-$0.059 & $-$ 0.044 & 0.023 & 0.063\\ [2.5ex]
 \hline\\
\end{tabular}
\end{table*}

\subsubsection{Bader$-$effective Charge Analysis}
In order to quantify the bonding interactions between constituents and estimate the electron distribution in the participating atoms, we have calculated the Bader's effective charge (BC) using the Bader's atom$-$in$-$molecule (AIM) concept by topological analysis of charge density. \cite{bader1990international, henkelman2006fast}. The AIM theory provides a definition for characteristics of chemical bonding that gives numerical values for the bond strength. The zero flux surfaces in the charge density are used to define each basin belonging to a particular atom. The charge associated to this atom is then obtained by integrating the charge density over the entire basin. The BC values of participating atoms in $M$(BH$_4$)$_2$(NH$_3$)$_2$ ($M$ = Ti, Mn, Fe, Cu and Zn) systems are calculated and tabulated in Table~\ref{BEC}.

In Ti based TMABHs, the BC values at the Ti site is 1.37 electrons and are shared with its neighbouring constituents since the Ti in this system has 2$+$ oxidation states. The BC values for B and H$^{-\delta}$ are around $+$1.5 e and $-$0.5 e, respectively. This suggests that around $+$1.5 e are transferred from B to H$^{-\delta}$. On the other hand, the BC value for N and H$^{+\delta}$ are $-$1.4 e and $+$0.5 e, respectively, indicating that the N ions draw charges from the neighbouring H due to its high electronagaitvity. The above results indicate the presence of partial ionicity between B$-$H$^{-\delta}$ as well as N$-$H$^{+\delta}$. The H$^{-\delta}$ and H$^{+\delta}$ BC values in all these TMABHs clearly show the presence of amphoteric hydrogen where H$^{+\delta}$ donated $+$0.523 electrons to the host lattice and H$^{-\delta}$ accepted $-$0.571 electrons from the neighbouring constituents.

\subsection{Decomposition mechanism of NH$_3$ in TMABHs systems} \label{DM}
Designing new hydrogen storage materials with favourable hydrogenation and dehydrogenation properties is one of the key challenges to use the hydrogen as an alternative energy carrier for mobile as well as stationary applications. Nakamori \textit{et al.}\cite{nakamori2006correlation} described the correlation between hydrogen desorption temperature and Pauling electronegativity ($\chi_{P}$) of the cation in metal borohydrides. The metal borohydrides with metal cation having low $\chi_{P}$ tend to release hydrogen at high temperature and in contrast cation with high $\chi_{P}$ tend to release diborane (B$_2$H$_6$) at low temperature. \cite{li2007materials} In order to improve the thermal stability of metal borohydrides, NH$_3$ ligands are used to saturate and immobilize the metal cations with formation of ammine metal borohydrides (AMBs, [$M$(NH$_3$)$_x$][BH$_4$]$_y$). Even though the AMBs show improved dehydrogenation properties and small amount of B$_2$H$_6$ release compared to that of metal borohydrides, many of these AMBs tend to release ammonia during thermal decomposition. So, in order to have deeper insight into the decomposition mechanism of NH$_3$ in TMABHs systems we have used reaction path calculation using the nudged elastic band method. 

\begin{table}[h]
\caption{The calculated NH$_3$ formation energy (E$_{vac}$ in eV), diffusion barrier for ammonia migration (E$_b$ in eV) and activation energies for NH$_3$ (E$_a$ in KJ mole$^{-1}$), hydrogen site energy for H closed to B (E$_{H^{-\delta}}$ in eV) and H close to N (E$_{H^{+\delta}}$ in eV) in $M$(NH$_3$)$_2$(BH$_4$)$_2$ $M$ = Ti, Mn, and Zn.}
\label{AEs}
\begin{tabular}{llllll}
\hline \hline\\ [-1.5ex]
Compound & E$_{vac}$  & E$_b$ & E$_a$ & E$_{H^{-\delta}}$ & E$_{H^{+\delta}}$ \\ 
\hline\hline \\ [1.5ex]
Ti(NH$_3$)$_2$(BH$_4$)$_2$ & 1.61 & 1.01 & 2.62 & 0.61 & 0.01\\ [1.5ex] 
Mn(NH$_3$)$_2$(BH$_4$)$_2$ & 1.60 & 1.47 & 3.07 & 1.56 & 0.57 \\ [1.5ex]
Zn(NH$_3$)$_2$(BH$_4$)$_2$ & 1.44 & 0.87 & 2.31 & 3.2 & 2.15\\ [1.5ex]
\hline 
\end{tabular}
\end{table}

In order to  understand the microscopic dehydrogenation mechanism of $M$(NH$_3$)$_2$(BH$_4$)$_2$ ($M$ = Ti, Mn and Zn), we have studied the NH$_3$ vacancy formation energy and also its diffusion barrier. The NH$_3$ vacancy formation energy was estimated using the following equation: 
\begin{equation}
E_{vac} = E(TMABHs-NH_3)+E(NH_3)-E_{total}
\end{equation}
Where E$_{total}$ is the total energy of the TMABHs; E(NH$_3$) represents the energy of an isolated NH$_3$ molecule; E(TMABHs$-$NH$_3$) is the total energy of the TMABHs after the removal of NH$_3$ structural sub-unit. As shown in Table~\ref{AEs}, the calculated NH$_3$ vacancy formation energies are 1.61, 1.60 and 1.44 eV for $M$(NH$_3$)$_2$(BH$_4$)$_2$, $M$ = Ti, Mn and Zn, respectively. The relatively high NH$_3$ removal energies indicates that the vacancy formation of NH$_3$ are thermodynamically unfavourable at low temperatures. 

The diffusion of NH$_3$ is considered as the migration of an NH$_3$ unit attached to a metal site to a nearby NH$_3$ vacancy site. The diffusion barrier is defined as the energy difference between the saddle point and the ground state. The energy barrier of ammonia diffusion is calculated to be 1.01, 1.47, and 0.87 eV for $M$(NH$_3$)$_2$(BH$_4$)$_2$ with $M$ = Ti, Mn, and Zn, respectively. Chen \textit{et al.} \cite{chen2017first} reported the ammonia diffusion barrier for Mg(BH$_4$)$_2$ $\cdot$ 2NH$_3$ and the corresponding value is 0.26 eV. Compared to Mg(BH$_4$)$_2$ $\cdot$ 2NH$_3$, the relatively high value of ammonia diffusion barrier for 3\textit{d} transition metal ammine borohydrides considered in the present study indicate that the  mobility of ammonia is low in $M$(NH$_3$)$_2$(BH$_4$)$_2$. Moreover, the relatively higher activation barrier of NH$_3$ release in TMABHs compared with Mg based metal ammine borohydrides suggest that the ammonia release need higher temperature. We have also calculated hydrogen site energy for hydrogen associated with NH$_3$ as well as BH$_4$ structural sub-units and those value for the representative system are given Table ~\ref{AEs}. The electronegativity value of Ti, Mn, and Zn are lower than that of Mg and hence they donate less charge to NH$_3$ and BH$_4$ structural sub-units and hence from this observation one can conclude that the TMABHs system have high diffusion barrier for NH$_3$, which limits the transport of ammonia, therefore improve the dehydrogenation.   

\begin{figure*}
 \centering
 \includegraphics[width=\linewidth]{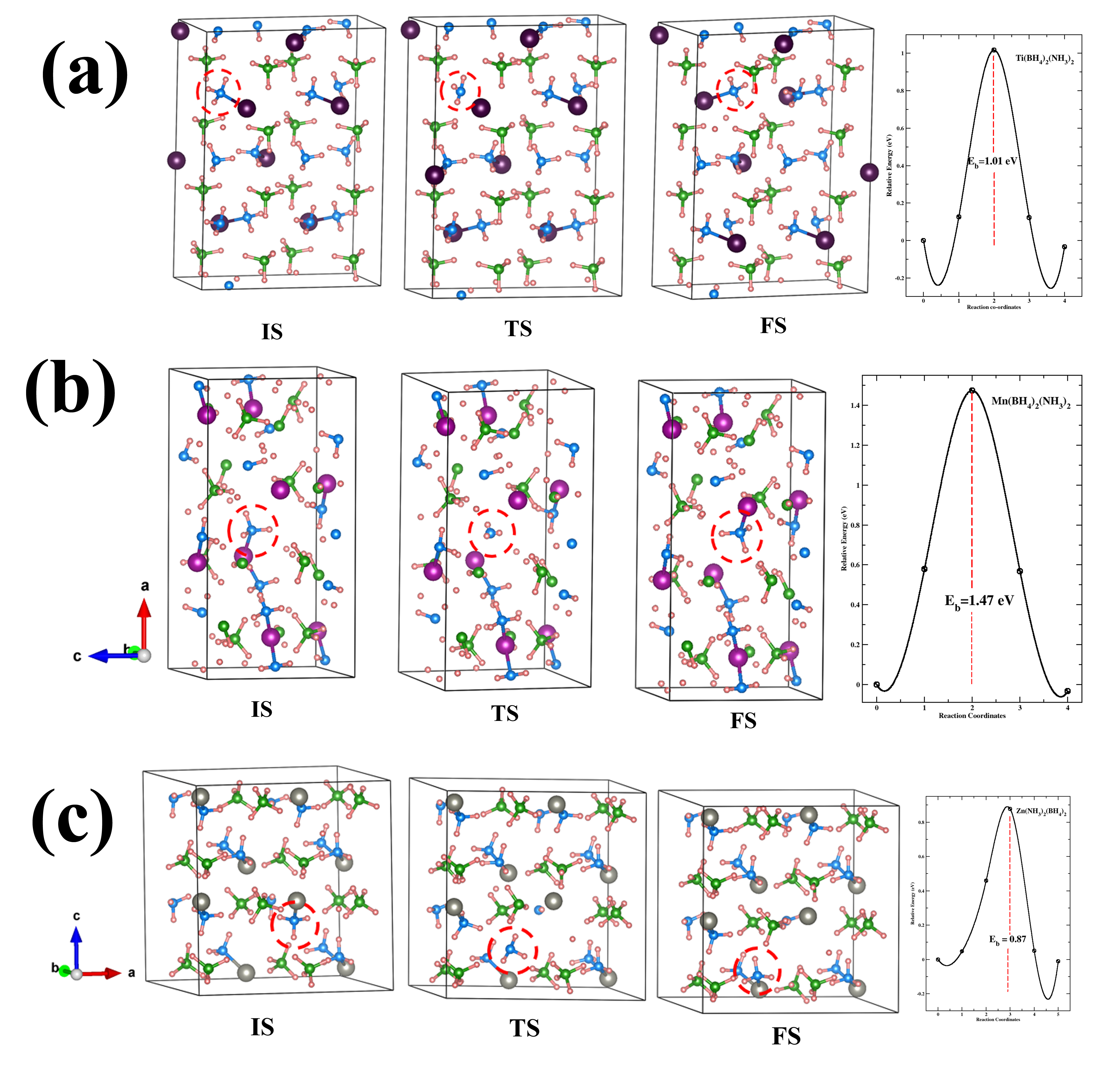}
 \caption{The calculated minimum energy path, initial (IS), transition (TS) and final (FS) geometric structure of NH$_3$ diffusion in $M$(NH$_3$)$_2$(BH$_4$)$_2$ M =Ti (a), M= Mn (b), and M= Zn (c), respectively. The brown, purple, grey, blue, green and pale red colors represent Ti, Mn, Zn, N, B, and H, atoms, respectively.}
 \label{neb}
\end{figure*}

The activation energy (E$_a$) for self diffusion of ammonia can be obtained by combining the calculated vacancy formation energy of NH$_3$ with the corresponding diffusion barrier. As summarized in Table~\ref{AEs} for $M$(NH$_3$)$_2$(BH$_4$)$_2$ ($M$ = Ti, Mn and Zn), the calculated activation energy of NH$_3$ diffusion are 2.62, 3.07 and 2.31 eV, for $M$ = Ti, Mn, and Zn, respectively. Compared with the diffusion barrier energy, the formation energy of NH$_3$ is higher and hence one would except that the amount of ammonia release will be small in these compounds during the decomposition process. As a consequence of this, it is expected that there will be low concentration of NH$_3$ vacancy in these systems.   

Welchman \textit{et al.} \cite{welchman2017decomposition} stated that,  TMABHs with $M$ having low$-$ $\chi_P$ (i.e. $\chi_P$ $\leq$ 1) value will have NH$_3$ weakly bound to the metal and as a result it will decomposes easily and release the NH$_3$. On the other hand, if $M$ is having mid$-$ $\chi_P$ (1 to less than 1.6) values then the release of NH$_3$ becomes less favourable and such metal based ammine borohydrides release H$_2$ directly during the decomposition which becomes more favourable when the $\chi_P$ value of $M$ is higher. Finally, for $\chi_P$ $\geq$ 1.6, the production of B$_2$H$_6$ or B$_2$H$_7$ becomes more favourable than direct H$_2$ release resulting in indirect release of H$_2$. In our system also we could except above stated results when we do thermal decomposition. It may be noted that both Ti and Mn have mid$-$ $\chi_P$ values and hence the release of NH$_3$ becomes less favourable and these TMABHs tend to direct release of H$_2$ during decomposition. Similarly, the Zn is having higher $\chi_P$ value than Ti and Mn and hence the release of NH$_3$ is relatively less favourable in Zn based TMABHs compared to Ti and Mn based TMABHs systems.

It is well known that, ammonia is consider as an attractive energy carrier to store hydrogen in liquid form. However, the toxicity and vapour pressure of liquid ammonia make it undesirable for direct use in mobile applications, mainly because of the potential risk of accidents where ammonia is released. Moreover, due to the technical challenges of securing the end user against contact with liquid ammonia during refilling and while performing periodical maintenance hinder its wide scale use.~\cite{duijm2005safety, eggmann2001kirk, thomas2006potential} It is to be noted that, transportation of liquid ammonia in closed systems to decentralized sites for production and regeneration of metal ammine salts can minimize the overall cost and be scaled up while maintaining safety. However, some additional steps need to be taken to minimize the risks involved. 

The idea of using ammonia as a hydrogen carrier has been promoted by the further development of safe storage of ammonia in solid form by binding it in metal ammine complexes.~\cite{christensen2005metal} The solid storage of ammonia solves the safety issues of driving with ammonia in liquid form under pressure. At the same time, the volumetric hydrogen density is high for the metal ammines compared to that of liquid ammonia and metal hydrides. The well known metal ammine complexes are Mg(NH$_3$)$_6$Cl$_2$ and Ca(NH$_3$)$_8$Cl$_8$ were selected due to the low vapour pressure at room temperature and high gravimetric (around 9.19 wt.\%) and volumetric (around 109 (g H$_2$) L$^{-1}$) hydrogen density. Among these systems Ca(NH$_3$)$_8$Cl$_8$ has shown promise for mobile applications and the release of ammonia is achieved at lower temperatures than that of Mg(NH$_3$)$_6$Cl$_2$. ~\cite{sorensen2008indirect} As suggested by the above studies, the TMABHs considered in  the present study also having high gravimetric capacity (10.92 wt \% to 12.98 wt\%) and high volumetric capacity ( 127.08 g H$_2$ L$^{-1}$ to 157.08 g H$_2$ L$^{-1}$) of hydrogen and hence we suggest that these materials are more suitable for solid storage of ammonia.

\section{Conclusion}
Even though the metal based borohydrides with addition of ammoniates have wide varieties of both experimental and theoretical results, the ground state crystal structure of $M$(BH$_4$)$_2$(NH$_3$)$_2$ ($M$ = Sc, Ti, V, Cr, Fe, Co, Ni, and Cu) is yet to be predicted. Using the state-of-the art density functional calculations we have predicted the ground state structure and the unit cell parameters of $M$(BH$_4$)$_2$(NH$_3$)$_2$. Also, we have studied the phase stability, electronic structure, magnetic properties, chemical bonding analysis, as well as decomposition mechanism and the conclusions obtained from these analyses are as follows:

$\bullet$ We have optimized the ground state crystal structure of experimentally known Zn(BH$_4$)$_2$(NH$_3$)$_2$ using various exchange correlation functionals and found that GGA overestimate the equilibrium volume by $-$3.33 \% and the optPBE$-$vdW functional underestimate it by 0.31\% so we conclude that one should account van der Waals interaction into the calculation to correctly predict the structural properties of TMBAHs. From the detailed analysis using various vdW functionals including optB86b$-$vdW functional done by earlier theoretical studies\cite{chen2018decomposition}, we have found that optPBE$-$vdW functional is the appropriate exchange correlation functional to reliably predict the equilibrium structural properties of TMABHs. Hence, we have used optPBE$-$vdW functional for predicting the equilibrium structural parameters for experimentally not yet explored TMABHs compounds considered in the present study. 

$\bullet$ We have predicted the ground state crystal structures of $M$(BH$_4$)$_2$(NH$_3$)$_2$ ($M$ = Sc, Ti, V, Cr, Mn, Fe, Co, Ni, and Cu) using the spin-polarised calculation with optPBE$-$vdW functional and their lattice parameters are tabulated and compared with available experimental results. The spin-polarised calculation of TMABHs attains antiferromagnetic ordering as lowest energy configuration for all these systems. Furthermore, magnetic TMABHs studied here obey the Slater$-$Paulings rule and it has been found that these systems have high value of magnetic moment with high spin states at the transition metal site. 

$\bullet$ We have calculated the reaction enthalpy of formation for AMABHs and TMABHs using three different reaction paths and found that all these compounds are experimentally feasible to synthesis using some of these reaction paths. From the calculated reaction enthalpy of formation we have concluded that Fe, Co, Ni, and Cu based TMABHs are feasible to synthesis using all these three reaction paths. 

$\bullet$ The calculated total density of states show a finite gap between the valence band maxima and conduction band minima for all these TMABHs. The partial DOS analysis shows that there is an ionic bonding between the metal atom and two structural sub-units (i.e. BH$_4$ and NH$_3$). The bonding between B$-$H$^{-\delta}$ and N$-$H$^{+\delta}$ enunciate the presence of covalent character between them with substantial ionicity. Moreover, our detailed analyses show that the H closer to B has negative oxidation state and that closer to N has the positive oxidation states confirming the presence of amphoteric hydrogen in TMABHs systems. 

$\bullet$ The amphtoric behaviour hydrogen in TMABHs where further substantiated using charge density distribution, electron localization function, Born effective charge and Bader charge anaylses. These analyses also reveal that TMABHs has ionic bonding between the metal atom and the two structural sub-units. Whereas the interaction between N$-$H$^{+\delta}$ and B$-$H$^{+\delta}$ has iono$-$covalent nature. The bonding nature between the constituents in all these systems point to the fact that the TMABHs have iono$-$covalent bonding.  

$\bullet$ From the decomposition mechanism study of TMABHs we found that the vacancy formation energy and diffusion barrier of NH$_3$ are relatively high and those findings indicating that the release of NH$_3$ happens at high temperatures whereas the direct H$_2$ release happen at low temperatures. 

With the above insights, we conclude that the predicted TMABHs system are more liable materials to be used for the practical hydrogen storage applications. We hope that the present study motivates further research in this direction by identifying transition metal based ammine borohydrides with amphoteric hydrogen for  practical hydrogen storage applications.

\section*{Acknowledgements}
The authors are grateful to the SCANMAT Centre, Central University of Tamil Nadu, Thiruvarur for providing the computer time at SCANMAT supercomputing facility. The authors are grateful to the Research Council of Norway for providing the computer time (under the project number NN2875k) at the Norwegian supercomputing facility.

\section{Computational Details}
The total energy calculations have been performed according to the projector augmented wave method (PAW)~\cite{blochl1994projector, kresse1999ultrasoft} as implemented in the Vienna \textit{ab$-$initio} simulations package (VASP).~\cite{kresse1993ab} The calculations were made by utilizing density functional theory, employing the generalized gradient$-$approximation (GGA) functional of Perdew \textit{et al.}~\cite{perdew1996generalized}  All the calculation were carried out with a 300\,eV plane wave cut$-$off and this energy cut$-$off value is found to be reliable to predict structural and the chemical bonding behaviour of similar systems earlier.~\cite{ravindran2006modeling} The structural optimizations were performed using force and stress minimization. For the structural optimization the \textbf{k}$-$points were generated using the Monkhorst pack method with the grid of $4\times3\times4$ for the monoclinic structure experimentally found for Zn(BH$_4$)$_2$(NH$_3$)$_2$ and similar \textbf{k}$-$point density were used for other compounds considered in the present study.

\paragraph{} To account the weak van der Waals(vdW) interaction between molecule$-$like structural sub-units, we have used various vdW$-$corrected density functional theory (DFT) methods implemented in the VASP package.~\cite{dion2004van,roman2009efficient,klimevs2009chemical,lee2010higher,klimevs2011van,thonhauser2007van}  
Among the various vdW functionals considered for the present study we found that the optPBE$-$vdW functional~\cite{thonhauser2007van} give reliable structural parameters and hence it is used for most of our analysis. The total energy was calculated as a function of volume using force as well as stress minimization to estimate the equilibrium volume and also check the dispersion behaviour of total energy vs. volume curves. To find the ground state magnetic ordering we have performed total energy calculations for non$-$magnetic (NM), ferromagnetic (FM), and various antiferromagnetic (AFM) configurations.~\cite{patra2016electronic}

\paragraph{} In order to quantify the charge at the various atomic sites we have adapted various charge partitioning schemes. Among them the Born effective charges are obtained for the ground$-$state structure in the equilibrium volume using modern theory of polarization with Berry phase approach.~\cite{klaveness2007structure} The Bader effective charges were obtained from the fine grid charge density generated from VASP calculations using the topological analysis of charge density with the help of Bader's atoms$-$in$-$molecule method.~\cite{yang2012ab}

\paragraph{}Computational methods for calculating minimum energy paths (MEP) are used in the field of theoretical chemistry, physics, and materials science. The MEP describes the reaction mechanism, and in thermal systems, the energy barrier along the path can be used to calculate the reaction rate. There are different approaches adapted to determine the MEP. We have restricted our investigation to nudged elastic band method in which the initial and final states are known.~\cite{henkelman2000climbing, henkelman2000improved, berne1998classical} 

\paragraph{}In order to get the deeper insight about the decomposition mechanism of NH$_{3}$ in TMABHs systems, the NH$_{3}$ diffusion barriers were estimated by using nudged elastic band (NEB) method~\cite{sheppard2012generalized, sheppard2011paths, sheppard2008optimization, jonsson1998nudged, henkelman2000improved}. The NH$_{3}$ vacancy was created by removing one NH$_{3}$ molecule from the ground state crystal structure of TMABHs using appropriate super cell. We have created a 2$\times$1$\times$2 supercell for Zn(NH$_{3}$)$_{2}$(BH$_{4}$)$_{2}$ and then we have removed one NH$_{3}$ molecule from the Zn based TMABHs systems. Similar, approach was followed for other TMABHs systems considered in the present study. 

\subsection{Theoretical models to account for van der Waals interactions}
The London dispersion interaction plays an important role for contributing to the binding of molecules on surfaces, biomolecules, molecular$-$like crystals etc. In order to account such dispersion interactions and also hydrogen bonds present in the systems, considered in the present study we have accounted the van der Waals interaction in to the calculations. The non$-$local van der Waals density functional (vdW$-$DF) of Dion \textit{et al.}~\cite{dion2004van} is a promising method to account the dispersion bond present in the systems. The van der Waals functional proposed by Dion \textit{et al.} for the exchange correlation term is

\begin{equation} \label{eqn:1}
E_{xc} = E^{GGA}_{x} + E_{c}^{LDA}+ E_{c}^{nl}
\end{equation}

where the exchange energy, E$_{x}^{GGA}$ is obtained from the GGA functional proposed by Perdew$-$Burke$-$Ernzerhof (PBE)~\cite{perdew1996generalized} and the correlation energy, E$_{c}^{LDA}$, is obtained from the local density approximation. The term E$_{c}^{nl}$ is the non$-$local energy term which accounts approximately for the non$-$local electron correlation effects. Rom{\'a}n$-$P{\'e}rez and Soler ~\cite{roman2009efficient} have replaced the double spatial integral into fast Fourier transforms, this speeds up the computational time.

The Tkatchenkoa Scheffler (TS) dispersion correction method uses fixed neutral atoms as a reference to estimate the effective volumes of atoms$-$in$-$molecule(AIM) to calibrate their ploarizabilities. However, the dispersion coefficients obtained from this method fails to describe the appropriate ground state structure and the energetics for ionic solids. This problem can be solved by replacing it with the conventional Hirshfeld partitioning used to compute properties of interacting atoms by the iterative scheme proposed by Bultinck. \cite{tkatchenko2009accurate, tkatchenko2012accurate, bucko2013improved, buvcko2014extending, buvcko2016many} In this iterative Hirshfeld algorithm, the neutral reference atoms are replaced by ions with fractional charges obtained from AIM approach in a single iterative procedure. This algorithm is initialized with a pro$-$molecular density defined by non$-$interacting neutral atoms. The iterative procedure then runs in the following steps:\\
1. The Hirshfeld weight function for the step i is computed as 
\begin{equation} \label{eqn:2}
\omega^i_A(r) = n^i_A(r)/\bigg(\sum_B n^i_B(r)\bigg)
\end{equation}
where the sum extends over all atoms in the systems\\
2. The number of electrons per atom is determined using 
\begin{equation}  \label{eqn:3}
N^{i+1}_A = N^i_A + \int[n^i_A(r)-\omega^i_A(r)n(r)]d^3r
\end{equation}
3. New reference charge densities are computed using 
\begin{equation}  \label{eqn:4}
n^{i+1}_A(r) = n^{lint(N^i_A)}(r)[uint(N^i_A)-N^i_A]+n^{uint(N^i_A)}(r)[N^i_A - lint(N^i_A)]
\end{equation}
where lint($x$) express the integer part of $x$ and uint($x$)=lint($x$)+1

The above mentioned steps ~\ref{eqn:2} to ~\ref{eqn:3} are iterated until the difference in the electronic populations between two subsequent steps ($\Delta^i_A = |N^i_A - N^{i+1}_A|$) are less then a predefined threshold for all atoms. The converged iterative Hirshfeld weights ($\omega^i_A$) are then used to define the AIM properties needed to evaluate dispersion energy.

Our previous studies show that the optPBE$-$vdW functional~\cite{kiruthika2019amphoteric} is appropriate vdW$-$correction to be included in the DFT method to accurately describe the structural parameters of similar systems and hence we have adopted that method also to predict the ground state structural properties. Also, for Zn(BH$_4$)$_2$(NH$_3$)$_2$ there is a recent report~\cite{chen2018decomposition} where they have used optB8b$-$vdW~\cite{klimevs2009chemical} functional to predict its structural properties and hence in order to reproduce their results we have considered optB8b$-$vdW functional also for calculating the equilibrium structural parameter for Zn(BH$_4$)$_2$(NH$_3$)$_2$.

\medskip
\textbf{Supporting Information} \par 
Supporting Information is available from the Wiley Online Library or from the author.

\medskip
\textbf{Acknowledgements} \par 
Please insert your acknowledgements here

\medskip

%
\bibliographystyle{MSP}
\bibliography{MSP-template}





\end{document}